\documentclass[twocolumn,times,tighten,trackchanges]{aastex62}
\usepackage{natbib}
\usepackage{epsfig}
\usepackage{wrapfig}
\usepackage{graphicx}
\usepackage{floatflt}
\usepackage{ulem}
\usepackage{natbib}
\usepackage{epsfig}

\submitjournal{Astrophysical Journal}
\shorttitle{Kinematic Properties of Young Brown Dwarfs}
\shortauthors{Riedel, DiTomasso et al.}

\begin{document}
\title{Radial Velocities, Space Motions, and Nearby Young Moving Group Memberships of Eleven Candidate Young Brown Dwarfs}
\footnote{Data presented herein were obtained at the W.M. Keck Observatory, which is operated as a scientific partnership among the California Institute of Technology, the University of California and the National Aeronautics and Space Administration. The Observatory was made possible by the generous financial support of the W.M. Keck Foundation.}

\author{Adric R. Riedel}
\affiliation{Space Telescope Science Institute, Baltimore, MD 21218, USA}
\affiliation{Department of Astronomy, California Institute of Technology, Pasadena, CA 91125, USA}
\affiliation{Department of Engineering Science and Physics, College of Staten Island, City University of New York, Staten Island, NY 10314, USA}
\affiliation{Department of Astrophysics, American Museum of Natural History, New York, NY 10024, USA}
\affiliation{Department of Physics and Astronomy, Hunter College, City University of New York, New York, NY 10065, USA}

\author{Victoria DiTomasso}
\affiliation{Leibniz-Institute for Astrophysics Potsdam (AIP), An der Sternwarte 16, D-14482 Potsdam, Germany}
\affiliation{Department of Astrophysics, American Museum of Natural History, New York, NY 10024, USA}
\affiliation{Department of Physics and Astronomy, Hunter College, City University of New York, New York, NY 10065, USA}

\author{Emily L. Rice}
\affiliation{Department of Physics and Astronomy, College of Staten Island, City University of New York, Staten Island, NY 10314, USA}
\affiliation{Department of Astrophysics, American Museum of Natural History, New York, NY 10024, USA}
\affiliation{Physics Program, The Graduate Center, City University of New York, 365 Fifth Ave, New York, NY 10016, USA}

\author{Munazza K. Alam}
\affiliation{Department of Astronomy, Harvard University, Cambridge, MA 02138, USA}
\affiliation{Department of Astrophysics, American Museum of Natural History, New York, NY 10024, USA}
\affiliation{Department of Physics and Astronomy, Hunter College, City University of New York, New York, NY 10065, USA}

\author{Ellianna Abrahams}
\affiliation{Department of Astronomy, University of California at Berkeley, Berkeley, CA 94720, USA}
\affiliation{Department of Astrophysics, American Museum of Natural History, New York, NY 10024, USA}
\affiliation{Department of Physics, City College of New York, City University of New York, New York, NY 10031, USA}

\author{James Crook}
\affiliation{Department of Astrophysics, American Museum of Natural History, New York, NY 10024, USA}
\affiliation{Hunter College High School, 71 East 94th Street, New York, NY 10128, USA}
\affiliation{Physics and Astronomy Department, University of California Los Angeles, 430 Portola Plaza, Los Angeles, CA 90095, USA}

\author{Kelle L. Cruz}
\affiliation{Department of Astrophysics, American Museum of Natural History, New York, NY 10024, USA}
\affiliation{Department of Physics and Astronomy, Hunter College, City University of New York, New York, NY 10065, USA}
\affiliation{Physics Program, The Graduate Center, City University of New York, 365 Fifth Ave, New York, NY 10016, USA}

\author{Jacqueline K. Faherty}
\affiliation{Department of Astrophysics, American Museum of Natural History, New York, NY 10024, USA}
\affiliation{Department of Terrestrial Magnetism, Carnegie Institution of Washington, DC 20015, USA}

\begin{abstract}
We present new radial velocity (RV) measurements for 11 candidate young very-low-mass stars and brown dwarfs, with spectral types from M7 to L7. Candidate young objects were identified by features indicative of low surface gravity in their optical and/or near-infrared spectra. RV measurements are derived from high resolution (R=$\lambda$/$\Delta\lambda$=20,000)~$J$ band spectra taken with NIRSPEC at the Keck Observatory. We combine RVs with proper motions and trigonometric distances to calculate three-dimensional space positions and motions and to evaluate membership probabilities for nearby young moving groups (NYMGs). We propose 2MASS~J00452143+1634446 (L2$\beta$, $J$=13.06) as an RV standard given the precision and stability of measurements from three different studies. We test the precision and accuracy of our RV measurements as a function of spectral type of the comparison object, finding that RV results are essentially indistinguishable even with differences of $\pm$5 spectral subtypes. 
We also investigate the strengths of gravity-sensitive K~{\sc i} lines at 1.24--1.25 $\mu$m and evaluate their consistency with other age indicators. We confirm or re-confirm four brown dwarf members of NYMGs -- 2MASS~J00452143+1634446, WISE~J00470038+6803543, 2MASS~J01174748$-$3403258, and 2MASS~J19355595$-$2846343 -- and their previous age estimates. We identify one new brown dwarf member of the Carina-Near moving group,  2MASS~J21543454$-$1055308. The remaining objects do not appear to be members of any known NYMGs, despite their spectral signatures of youth. These results add to the growing number of very-low-mass objects exhibiting signatures of youth that lack likely membership in a known NYMG, thereby compounding the mystery regarding local, low-density star formation.
\end{abstract}


\section{\sc Introduction}
\label{sec:introduction}
Studying brown dwarfs is our gateway to constraining the formation and evolutionary histories of giant planets and their atmospheres. Brown dwarfs, especially young objects, can have masses and temperatures comparable to directly-imaged exoplanets \citep{Liu13}, but as free-floating objects rather than as stellar companions, they are more amenable to detailed study with current instrumentation. With the current generation of high contrast integral field spectrograph instruments such as Project 1640, GPI, and SPHERE \citep{Oppenheimer13,Macintosh08,Beuzit08} and soon JWST  \citep{Seager09}, the question of giant planet atmospheres and their formation is an increasing focus. 

Brown dwarfs do not achieve stable hydrogen fusion; therefore, they have no main sequence and no direct mass-luminosity relationship. Instead, brown dwarfs continually decrease in radius, temperature, and luminosity over time. It is thus difficult to tell the difference between brown dwarfs of different masses based on spectra alone; a young low-mass brown dwarf can have the same temperature as an old high-mass brown dwarf. Indeed, for many very-low-mass objects it is not possible to determine whether an object is a star or a brown dwarf without an estimate of the object's age. There are two ways to resolve this mass-age degeneracy: dynamical mass measurement \citep[e.g.,][]{Konopacky10,Dupuy14,Dupuy15}, which uses a combination of astrometry and spectroscopy to determine dynamical masses; and age measurements, which currently rely on spectroscopic and kinematic diagnostics. Dynamical masses require the brown dwarf to be in a close binary system, which is rare (2.5 $\rm ^{+8.6}_{-1.6}$\% of the population, \citealt{Blake10}), and a complete (or at least partial) orbit, which can require years to decades of astrometric monitoring. Precise age measurements for field-age and younger brown dwarfs (i.e., non-subdwarfs) require either a stellar companion with a reliable age constraint or membership in a nearby young moving group (NYMG), cluster, or star forming region where age constraints are then provided by the NYMG as a whole, typically based on age constraints determined using higher-mass members \citep[e.g.,][]{Zuckerman04}.

For young, single brown dwarfs, the most readily accessible method to estimate age is via kinematic membership in a NYMG. The NYMGs are, as their name implies, groups of stars and brown dwarfs moving together through space with similar space velocities. The assumption is that they formed together in a single star-forming event, with the same Galactic orbits as their natal molecular cloud. Though they are not gravitationally bound to each other in an open cluster, they are still young enough to shear from the Galactic potential and that chance encounters with disk stars have not completely obscured their shared trajectory. As such, determining the space velocity (and space position) of young objects is a powerful method of determining their potential membership in a nearby young moving group. NYMGs are sparse, containing perhaps a few hundred members spread out over thousands of cubic parsecs. Known groups include $\beta$ Pictoris \citep[$\sim$20 Myr,][]{Mamajek14}, Tucana-Horologium ($\sim$45 Myr, \citealt{Bell15}), Argus ($\sim$50 Myr, \citealt{BarNav99}) and AB Doradus \citep[$\sim$120 Myr,][]{Binks14,Bell15}. NYMGs are windows into the later stages of star and planetary system formation. At these ages, natal gas and dust are dissipated, removing extinction within the brown dwarf system, but brown dwarfs and very low mass stars are still physically enlarged compared to their field ($>$1 Gyr) equivalents. Thus they can exhibit spectral signatures of low surface gravity and potentially have different atmospheric cloud conditions and weather patterns \citep[e.g.,][]{Lew16}.

Probabilities of membership for individual objects in NYMGs are optimally calculated with complete spatial and velocity information, i.e., position, distance, proper motion, and radial velocity. While it is possible to determine memberships with only partial kinematics, \citet{Riedel17} demonstrates the importance of having better and more complete data. As shown in that paper, a brown dwarf can at best be given a 40\% probability of membership in $\beta$~Pictoris given only proper motion information; by that same token, the maximum probability rises to over 90\% with the addition of radial velocity information, even without a distance. Age constraints provided by NYMG membership can range from 5~Myr ($\epsilon$ Cham\ae leon, \citealt{Murphy13}) to 500~Myr ($\chi^{01}$ For, \citealt{Pohnl10}) with uncertainties of $\pm$10~Myr for TW Hydra \citep{Weinberger13} to $\pm$100~Myr for older groups.

The established memberships of NYMGs are deficient in low-mass members (mid-M dwarfs and later) relative to the field Initial Mass Function \citep[e.g.,][]{Jeffries12,Kraus14,Gagne17,Shkolnik17}. In order to complete the low-mass census of NYMGs, candidate young, very-low-mass objects are typically identified based on near-infrared (NIR) colors and low-resolution spectral features indicative of low surface gravity. Young very-low-mass objects are typically 1-2 magnitudes redder than the average NIR color for their spectral type \citep{Faherty12}. Spectra of these unusually red objects often exhibit spectroscopic signatures of low gravity, including weaker singly-ionized alkali metal lines, which is often taken to be a sign of youth \citep[e.g.,][]{Cruz09}. These objects are assumed to be young, with spectral type suffixes coarsely defined according to the divergence of gravity-sensitive spectral features from those of field (i.e., old) objects \citep{Cruz09,Allers13}. Finer age estimation based on spectral features alone is not currently possible; therefore, establishing membership in a NYMG is essential to providing age constraints for very-low-mass objects.

There are currently over 160 objects with spectral types M7 and later that have been identified as candidate members of nearby young moving groups. Prominent early examples included TWA 27 (2MASS J12073346$-$3932539, hereafter 2M1207$-$39) in TW~Hydra \citep{Gizis02}, 2MASS J01415823$-$4633574 (hereafter 2M0141$-$46, \citealt{Kirkpatrick06}) in Tucana-Horologium \citep{Gagne15b}, 2MASS J06085283$-$2753583 (hereafter 2M0608$-$27) in $\beta$~Pictoris (\citealt{Rice10}, but see also \citealt{Gagne14a} and \citealt{Faherty16}), and 2MASS J03552337+1133437 (hereafter 2M0355+11) in AB~Doradus \citep{Faherty13,Liu13}. These objects have frequently been used as comparison objects for newly discovered candidate young low-mass objects and even directly-imaged exoplanets \citep[e.g.,][]{Miles18, Greenbaum18, Crepp18}. There are additionally over 150 very-low-mass stars and brown dwarfs that display signatures of youth but lack complete kinematic information \citep[e.g.,][]{Gagne14a,Gagne14c,Gagne15a,Gagne15c,Faherty16}.

Of the spatial and kinematic data required for evaluating NYMG membership, radial velocity (RV) and parallax are arguably the most challenging measurements for intrinsically faint low mass targets. Multiple parallax programs \citep[e.g.][]{Gaia18,Faherty16,Liu16,Dieterich14,Marocco13,ZapOso14} are tackling the problem of parallaxes, which leaves radial velocities as the final important piece of the kinematic puzzle. Radial velocities for low mass objects necessitate either long exposure times to obtain sufficiently high resolution and signal to noise ratio spectra for RV measurements, even on the Keck 10-m telescope \citep[e.g.,][]{Blake10,Prato15}.
Further, absolute RV measurements are optimally calibrated against high-quality spectra of similar spectral type objects with existing RV measurements, which are only recently beginning to exist in large enough numbers to evaluate the dependence of measured RV on spectral type, signal to noise ratio (SNR), and other properties of the comparison object's spectrum.

In this paper we present new high resolution NIR spectroscopy, obtained with NIRSPEC on Keck~II, of 11 candidate young late-M and L dwarfs. We measure radial velocities to derive three-dimensional space positions and motions for the sample, and use them to determine membership, and therefore ages, in NYMGs. Unlike similar studies that use the $K$ band \citep{Blake10} or the $H$ band \citep{Faherty16}, we focus on the $J$ band, which contains numerous water absorption lines, prominent bandheads of FeH, and regions that are largely free of telluric absorption that can be used for cross-correlated RV measurements \citep{Prato15}. The $J$ band also contains strong alkali metal lines that are sensitive to surface gravity \citep[e.g.,][]{McLean07,Rice10}. The objects in our sample could be very low mass stars or brown dwarfs, depending on their ages, but we refer to them as brown dwarfs for the sake of simplicity.

In Section \ref{sec:sampleobsdr} we describe our sample of 11 M and L dwarfs, the NIRSPEC/Keck~II observations, and the data reduction procedure. In Section \ref{sec:analysis}, we describe the analysis and results, including RV measurements, the calculation of space positions and motions, and the five methods for estimating NYMG membership probabilities. We present notes on results for individual objects in Section~\ref{sec:notes}. In Section \ref{sec:discussion}, we discuss implications of our results for measuring RV of late-type objects and for evaluating various youth indicators. We present our conclusions in Section \ref{sec:conclusions}.

\section{\sc Sample, Observations, and Data Reduction}
\label{sec:sampleobsdr}
\subsection{Sample Selection}

\begin{figure}
\centering
\includegraphics[width=0.5\textwidth]{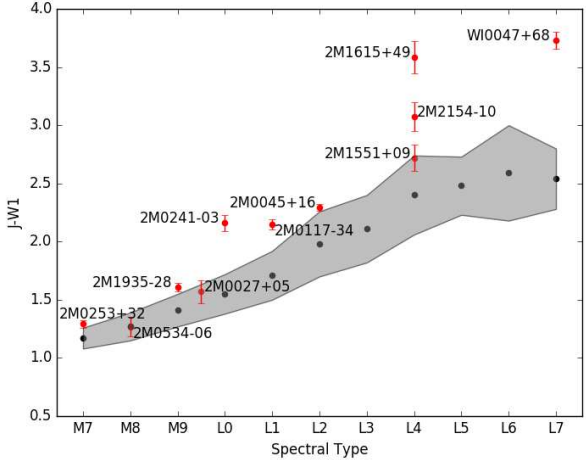}
\centering
\caption{Spectral type versus $J-W1$ color diagram for the 11 sample objects (red points) plotted with the average $J-W1$ colors (black points) and 1$\sigma$ spreads (gray shaded region) from \citet{Faherty16}. Eight of the 11 sample objects are more than 1$\sigma$ redder in $J-W1$ color than the average for their spectral type, especially the L dwarfs. The remaining three are more than 1$\sigma$ red in other color combinations.}
\label{fig:JW1}
\end{figure}

Our targets were selected from a sample of $\sim$M7 and later dwarfs identified as candidate young objects via their their classification as low-surface-gravity objects by \citet{Cruz09} using red-optical spectra and/or \citet{Allers13} using low-resolution, near-infrared spectra. All of the objects also have unusually red NIR colors for their spectral type (though not all in the specific $J-W1$ color shown in Figure~\ref{fig:JW1}). Eleven objects were observed during four half-nights in 2014 March and September; details of the observations are described in Section \ref{sec:observations} below and in Table~\ref{tab:observations}. 

There are only minor discrepancies between the optical spectra and the NIR spectroscopy seen here. The largest differences in spectral classification is 2MASS~J02411151$-$0326587 (hereafter 2M0241-03), which is an L0$\gamma$ object by optical spectral typing \citep{Faherty16} but an L1 VL-G by infrared spectral typing \citep{Allers13}; and with 2MASS~J02535980+3206373 (hereafter 2M0253+32), which was a M7$\beta$ by optical typing \citep{Faherty16} but was assigned as an M6 FLD-G in \citet{Allers13}.

All of these targets appear in \citet{Faherty16}, where eight of them were identified as having kinematics that suggested possible membership in multiple NYMGs or that could not be distinguished from field objects (``Ambiguous Member''). Two targets were determined by \citet{Faherty16} to be bona fide group members: 2MASS~J00452143+1634446 (hereafter 2M0045+16), identified as an Argus member by \citealt{Gagne14a}, and WISE~J004701.09+680352.2 (hereafter W0047+68), identified as an AB Doradus member by \citealt{Gizis15} and \citealt{Liu16}. 2MASS~01174748$-$3403258 (hereafter 2M0117$-$34) was listed as a high-likelihood member of Tucana-Horologium by both \citet{Faherty16} and \citet{Liu16}. \citet{Faherty16} presented RV measurements for three of our targets, one of which (2MASS~J00452143+1634446, hereafter 2M0045+16) was also previously measured by \citet{Blake10}. The other two were measured from low-quality spectra, motivating our decision to observe them again. In the time between our observations and this publication, \citealt{Gizis15} reported a radial velocity for W0047+68. These literature RV measurements are presented and compared to our RV results in Section~\ref{sec:RVs}.

\subsection{Observations}
\label{sec:observations}
Observations were made UT 2014 May 22 \& 24 and UT 2014 September 16 \& 17 using NIRSPEC, the cryogenic cross-dispersed echelle spectrometer on the Keck~II 10-m telescope at the W.M. Keck Observatory on Maunakea, Hawai'i \citep{McLean98,McLean00}. We used NIRSPEC's cross-dispersed echelle mode with the NIRSPEC-3 (N3) filter, which approximates standard $J$ band coverage (1.143--1.375~$\mu$m). In echelle mode, eight usable dispersion orders (65 to 58) are captured on the detector. Because the spectral interval captured by the detector is slightly smaller than the free spectral range in each order, there are small gaps, increasing with wavelength, in the total spectral coverage. The exact wavelength ranges for each dispersion order are listed in the headings in Section~\ref{sec:orders}. The slit width is three pixels (0\farcs432) for echelle observations. The resolving power in $J$ band is approximately R=$\lambda$/$\Delta\lambda$=20,000 (``high'' resolution) in echelle mode. Throughout the paper the high resolution $J$ band spectra are referred to by the number of the dispersion order, from 65~($\sim$1.17$\mu$m) -- 58~($\sim$1.31$\mu$m).

Observing methods follow those described in detail by \citet{McLean07} and \citet{Prato15}; the following is a brief summary and explanation of departures from those methods. Observations were made in pairs, nodding along the slit between each integration so that traces were separated by 7\arcsec~on the 12\arcsec-long slit. Due to a desire to avoid an intermittent quadrant in the slit-viewing camera, recent high resolution observations have used a smaller nod length. During these occasions the nod size was at least 2\arcsec~so that the dispersed traces would be well-separated on the slit. Integration time was 600~seconds per nod for all observations except for all four exposures of 2M0253+32 and four (of eight total) exposures of 2MASS~J05341594-0631397 (2M0534-06), which were 480~seconds per nod.

Total integration times per object are listed in Table~\ref{tab:observations}. A0~V stars were observed at an airmass very close to that of the target object (typically $<$0.1 airmass difference) to allow calibration for telluric absorption features. Arc lamp spectra were obtained at least once per night, and white-light spectra and corresponding dark frames were obtained for flat-fielding.

\begin{deluxetable*}{lllllrcrcll}
\tabletypesize{\scriptsize}
\tablewidth{0pt} 
\tablecaption{Observing Log \label{tab:observations} }
\tablehead{ 
\colhead{Object} &
\colhead{2MASS} &
\colhead{Optical} &
\colhead{NIR} &
\colhead{Sp. Type} &
\colhead{$\alpha$} &
\colhead{$\delta$} &
\colhead{$J$\tablenotemark{b}} &
\colhead{Int.\@~Time} &
\colhead{Average} &
\colhead{UT Date of}  \\
\colhead{Name\tablenotemark{a}} &
\colhead{ID}  &
\colhead{Sp.\@~Type} &
\colhead{Sp.\@~Type} &
\colhead{Ref} &
\colhead{J2000.0} &
\colhead{J2000.0} &
\colhead{mag} &
\colhead{seconds} &
\colhead{SNR} &
\colhead{Observation} }
\startdata
2M0253$+$32 & 02535980+3206373 & M7$\beta$  & M6 FLD-G & 1,2 & 02 53 59.70 & $+$32 06 37.0 & 13.62 & 1920 & 21 & 2014 September 16  \\
2M0534$-$06 & 05341594$-$0631397 & M8$\gamma$ & M8 VL-G & 2,1 & 05 34 15.94 & $-$06 31 39.7 & 16.05 & 4320 & 5.4 & 2014 September 17 \\
2M1935$-$28 & 19355595$-$2846343 & M9$\gamma$ & M9 VL-G & 2,1 & 19 35 55.96 & $-$28 46 34.4 & 13.95 & 3600 & 25 & 2014 May 22 \\
2M0027$+$05 & 00274197+0503417 & M9.5$\beta$& L0 INT-G & 2,1 & 00 27 41.97 & $+$05 03 41.7 & 16.19 & 4800 & 5.0 & 2014 September 16 \\
2M0241$-$03 & 02411151$-$0326587 & L0$\gamma$ & L1 VL-G & 3,1 & 02 41 11.50 & $-$03 26 58.0 & 15.80 & 4800 & 7.5 & 2014 September 16  \\
2M0117$-$34 & 01174748$-$3403258 & L1$\beta$ & L1 INT-G & 2,1 & 01 17 47.40 & $-$34 03 25.0 & 15.18 & 4800 & 10 & 2014 September 17 \\
2M0045$+$16 & 00452143+1634446 & L2$\beta$  & L2 VL-G & 3,1 & 00 45 21.43 & $+$16 34 44.6 & 13.06 & 1200 & 38 & 2014 September 16  \\
2M1551$+$09 & 15515237+0941148 & L4$\gamma$ & L4 VL-G & 2,1 & 15 51 52.37 & $+$09 41 14.8 & 16.32 & 7200 & 7.3 & 2014 May 24 \\
2M1615$+$49 & 16154255+4953211 & L4$\gamma$ & L3 VL-G & 2,1 & 16 15 42.50 & $+$49 53 21.0 & 16.79 & 7200 & 4.5 & 2014 May 22 \\
2M2154$-$10 & 21543454$-$1055308 & L4$\beta$ & L5$\gamma$ & 4,2 & 21 54 34.50 & $-$10 55 30.0 & 16.44 & 5400 & 4.3 & 2014 May 24  \\
W0047$+$68 & 00470038+6803543 & L7 ($\gamma$?) & L7.5 pec & 2,5 & 00 47 01.06 & $+$68 03 52.1 & 15.60 & 4800 & 6.9 & 2014 September 17 \\
\enddata
\tablenotetext{a}{2MASS, DENIS, and SDSS object names are truncated in subsequent tables and in the text.}
\tablenotetext{b}{From 2MASS All-Sky Point Source Catalog. } 
\tablerefs{Optical spectral types are those defined in \citet{Cruz09}, and near-infrared spectral types are on the scale defined in \citet{Allers13}. Spectral type suffixes indicate the strength of gravity-sensitive features, with $\beta$ is roughly equivalent to {\sc int-g} and $\gamma$ to {\sc vl-g}. Individual references are: (1) \citet{Allers13}, (2) \citet{Faherty16},  (3) \citet{Cruz09}, (4) \citet{Gagne14c}, (5) \citet{Gizis12}.}
\end{deluxetable*}

\subsection{Data Reduction}
\label{sec:data_reduction} 

All of the observed data were reduced with the REDSPEC IDL-based software package\footnote{\url{http://www2.keck.hawaii.edu/inst/nirspec/redspec.html}}, described in \citet{McLean03,McLean07}. The package performs standard bad pixel interpolation, dark subtraction, and flat-fielding as well as spatial rectification of curved spectra. Spectra are rectified and extracted in subtracted nod pairs so that the sky background and OH emission lines are removed. Spectra were extracted by summing over 7--15 rows dependent on seeing, then subtracted again to produce a positive spectrum with residual sky emission features removed. Most orders were wavelength calibrated with OH night sky lines, which were found to be highly stable and well-distributed across orders. Seven high resolution dispersion orders were reduced, covering orders 58--65 with the exception of order 60, where the OH night sky lines are blended with O$_2$ emission bands at 1.26--1.28~$\mu$m \citep{Rousselot00} making wavelength calibration and sky subtraction considerably more difficult. Each reduced spectrum was continuum normalized, and multiple nod pairs were averaged together to increase SNR. Spectra were shifted to the heliocentric reference frame using barycentric corrections calculated using JSkyCalc\footnote{\url{http://www.dartmouth.edu/~physics/labs/skycalc/flyer.html}}.

\subsection{Spectral Orders}
\label{sec:orders}
\begin{figure}
\centering
 \includegraphics[width=3.7in]{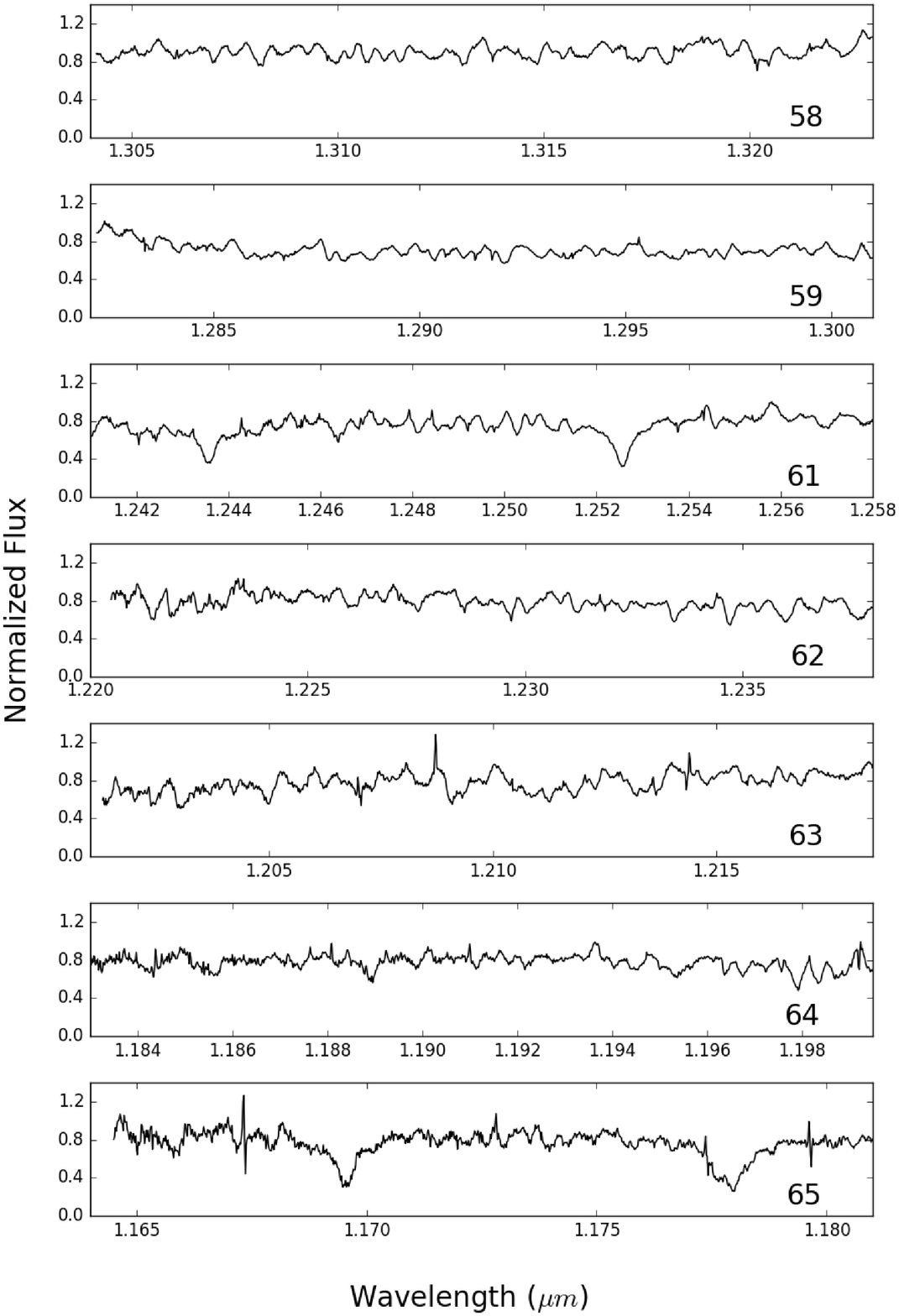}
\centering
\caption{NIRSPEC spectra for dispersion orders 58, 59, and 61 to 65 for the L2 object 2M0045+16. }
\label{fig:allorders}
\end{figure}

We present all reduced NIRSPEC dispersion orders for 2M0045+16 (L2$\beta$, $J$=13.06) in Figure~\ref{fig:allorders}. We summarize the relevant absorption features apparent in high resolution M and L dwarf spectra by NIRSPEC dispersion order below. More details can be found in \citet{McLean07} and \citet{Rice10}.\\

{\it Order 58 (1.30447 -- 1.32370 $\mu$m)} --- Al~{\sc i} doublet in the center of the order.\\

{\it Order 59 (1.28262 -- 1.30151 $\mu$m)} --- Weak Fe~{\sc i} lines.\\

{\it Order 60 (1.26137 -- 1.27999 $\mu$m)} --- Not reduced (see \ref{sec:data_reduction} above).\\

{\it Order 61 (1.24081 -- 1.25913 $\mu$m)} --- K~{\sc i} lines, higher SNR than order~65.\\

{\it Order 62 (1.22093 -- 1.23899 $\mu$m)} --- Used for RV measurements because of FeH and H$_2$O, strong well-spaced OH night sky lines, and weak telluric lines.\\

{\it Order 63 (1.20168 -- 1.21938 $\mu$m)} --- FeH and H$_2$O similar to order~62. \\

{\it Order 64 (1.18293 -- 1.20011 $\mu$m)} --- Weak Ti~{\sc i} and Mn~{\sc i} lines.\\

{\it Order 65 (1.16496 -- 1.18207 $\mu$m)} --- K~{\sc i} lines, typically lower SNR than in order~61 and blended with H$_2$O lines. \\

\begin{figure}
\centering
\includegraphics[width=3.7in]{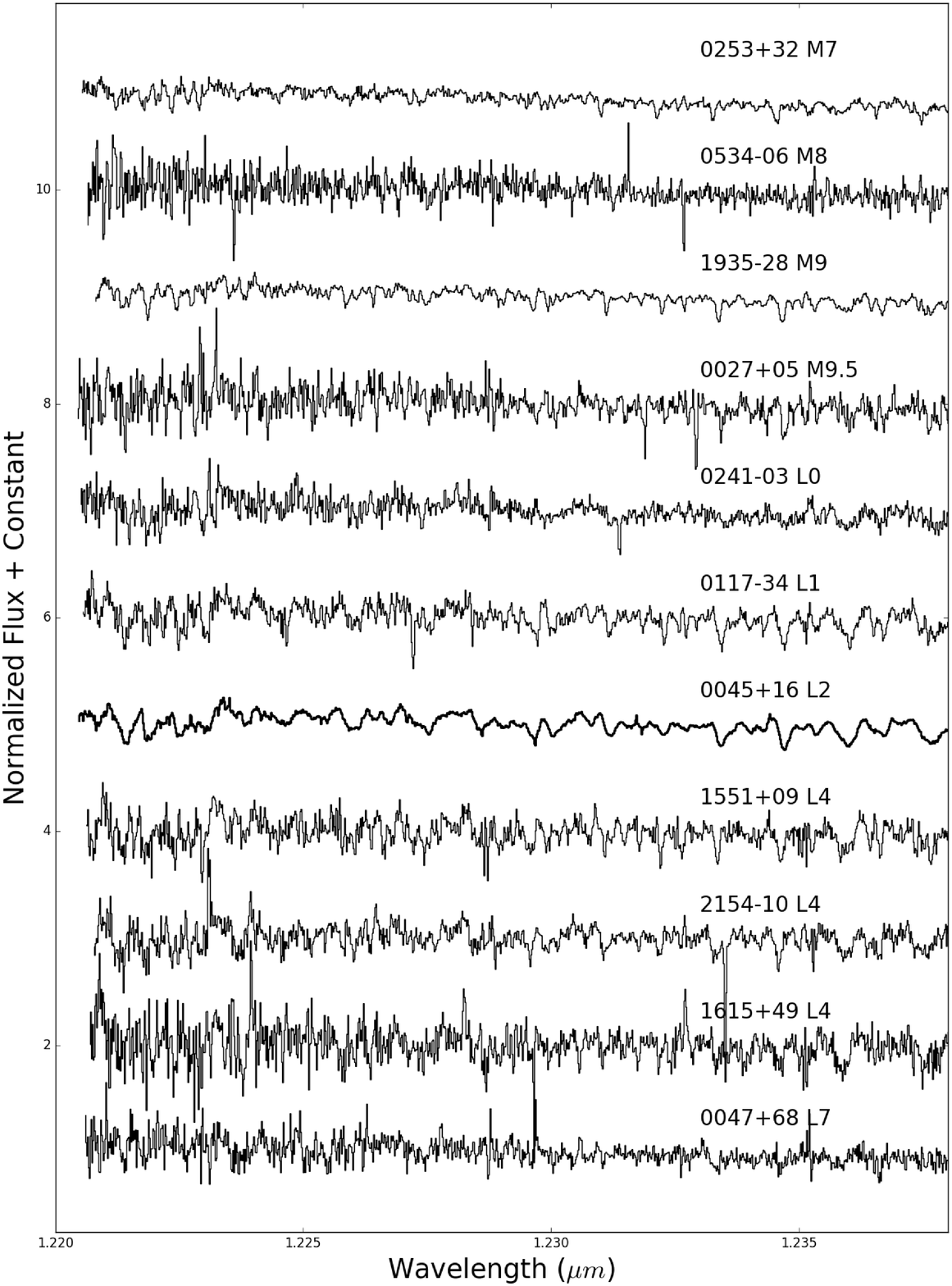}
\centering
\caption{NIRSPEC dispersion order 62 spectra (1.22093 -- 1.23899 $\mu$m) for 11 objects in the sample. }
\label{fig:order62}
\end{figure}

Following the work of the NIRSPEC Brown Dwarf Spectroscopic Survey \citep{Prato15} we use order 62 (1.221 -- 1.239 $\mu$m) for our radial velocity measurements. This wavelength regime is essentially free of telluric features, and M and L dwarf spectra contain numerous molecular absorption lines from FeH and H$_2$O that are ideal for cross-correlation techniques. Figure~\ref{fig:order62} presents the order 62 spectra for all 11 objects in the sample. Our targets have brightnesses between $J$=13 and $J$=17. Even with total integration times of 20 minutes to two hours (listed in Table~\ref{tab:observations}), the resulting spectra have average SNR in order~62 between 4 and 38, with a maximum SNR=38.2 for 2M0045+16 ($J$=13.06) and minimum SNR=4.3 for 2M2154$-$10 ($J$=16.44).

We also tested RV measurements using order 59, which is free of strong telluric absorption and was used by \citet{Prato15} for cross-correlating spectra of T dwarfs. For M and L dwarfs, the intrinsic atomic lines and molecular absorption lines at these wavelengths are weaker and the results were far less reliable. No order 59 results are presented in this paper.

\section{Analysis}
\label{sec:analysis}
\subsection{Radial Velocity Measurements}
\label{sec:RVs}
To measure radial velocity (RV), we use a custom cross-correlation code written in Python, first described in \citet{Faherty16}. The inputs are a heliocentric-corrected stellar spectrum (wavelength, flux, and uncertainty) and the spectrum of a comparison object with a previously measured RV, taken with the same instrumental setup to avoid systematics. For comparison spectra, we use objects with NIRSPEC Brown Dwarf Spectroscopic Survey (BDSS) spectra from \citet{McLean07}, \citet{Rice10}, and \citet{Prato15}. For the radial velocities of these standards, we use values reported in \citet{Rice10}, \citet{Blake10}, \citet{Chubak12}, and \citet{Prato15}, listed in Table \ref{tab:standards}.

\begin{deluxetable}{llccc}
\tablecaption{Comparison Stars\label{tab:standards}}
\tablehead{
\colhead{Name} &
\colhead{Spectral} &
\colhead{RV} &
\colhead{RV} &
\colhead{Average}
\\
\colhead{ } &
\colhead{Type} &
\colhead{km s$^{-1}$} &
\colhead{Ref.} &
\colhead{SNR}
} 
\startdata
GJ 406   & M6   &  19.321$\pm$0.145 & 1 & 326 \\
DENIS-P J1605-24 & M6 & $-$5$\pm$2 & 2 & 41 \\
SCH J1612$-$20 & M6.5	& $-$7$\pm$2 & 2 & 33\\
LP 402-58 & M7 & $-$3$\pm$2 & 2 & 65\\
UScoCTIO 130 & M7.5	& $-$7$\pm$2 & 2 & 23 \\
LP 412-31 & M8 & 42$\pm$2 & 2 & 118 \\
2MASS J1207$-$39AB & M8 & 8$\pm$2 & 2 & 38 \\
2MASS J0608$-$27 & M8.5 & 23$\pm$2 & 2 & 32 \\
2MASS 0140+27  & M8.5 &  9$\pm$2 & 3 & 87 \\
2MASS 0345+25  & L0   &  6$\pm$3 & 3 & 29 \\
2MASS 0746+20\tablenotemark{a} & L0.5 & 52.37$\pm$0.06 & 4 & 98 \\
2MASS 0208+25  & L1   & 20$\pm$2 & 3 & 14 \\
2MASS 2057$-$02  & L1.5 & $-$24.68$\pm$0.43 & 4 & 45 \\
2MASS 0015+35  & L2   & $-$37.35$\pm$0.16 & 4 & 49 \\
2MASS 2104$-$10  & L2.5 & $-$21.09$\pm$0.12 & 4 & 38\\
G 196-3B & L3 $\beta$ & $-$2$\pm$2 & 3 & 8 \\
2MASS 0036+18  & L3.5 & 19.02$\pm$0.15 & 4 & 86 \\
2MASS 2224$-$01  & L4.5 & $-$37.55$\pm$0.09 & 4 & 26 \\
2MASS 1507$-$16  & L5   & $-$39.85$\pm$0.05 & 4 & 76 \\
\enddata
\tablenotetext{a}{Binary}
\tablerefs{(1)\citet{Chubak12}, (2) \citet{Rice10}, (3) \citet{Prato15}, (4) \citet{Blake10} }


\end{deluxetable}

The target and standard spectra are read in and interpolated onto a log-normal spaced wavelength grid covering only the region where the spectra overlap, up-sampled in wavelength by a factor of 10. A third-order fit to the spectra is removed, taking out the large-scale structure of the spectra and leaving only the spectral lines. The code then re-samples each flux point of the target and comparison spectra from within the estimated noise on each, which we model as Gaussian random noise. The resulting re-sampled target spectrum is cross-correlated with the re-sampled radial velocity comparison spectrum, and the cross-correlation results are fit with a Gaussian+linear function to determine the velocity shift in pixels. The process of re-sampling the noise is repeated 1000 times, producing 1000 velocity shift measurements for each target and comparison object spectra pair, which we bin into a histogram. The histogram of velocity measurements is fit with a Gaussian function, the mean of which we adopt as the velocity shift, and the 1-$\sigma$ width of which we adopt as the uncertainty on the measurement, accounting for the noise of both spectra. The result then is converted from pixel shifts to velocity in km s$^{-1}$, and the known velocity of the comparison object is then subtracted to provide the actual heliocentric radial velocity of the target. The final uncertainty of each RV measurement is the combination of the uncertainty from the cross-correlation procedure and the uncertainty in the previously measured RV of the comparison object, added in quadrature. The latter dominates the precision of the results.

This cross-correlation technique is subject to comparison object-dependent systematic errors, including uncertainties on the wavelength solution of each observation and on the previously measured radial velocity of the comparison object. To test the accuracy of our RV results, we cross-correlate our targets with all 19 comparison objects in Table~\ref{tab:standards} using spectra from \citet{Prato15} and \citet{Rice10}, producing 19 individual RV measurements for each of our target objects. The mean, weighted by the calculated uncertainty, of these individual RV results constitutes the final calculated RV for each target object, as listed in Table \ref{tab:kinematicdata}. This procedure was used to calculate the RVs of eight out of eleven of our target objects.

The remaining three objects' RVs were calculated with slight modification to the routine described above. For these three targets, the cross-correlation process produced outlying pixel shifts that implied unrealistic velocities. For those three objects, W0047+68, 2M0534$-$06, and 2M0241$-$03, we only retain pixel shifts within plausible velocity ranges (between $\pm$500 or $\pm$100 re-sampled wavelength pixels, depending on object), and run the Monte Carlo iterations until 1000 pixel shifts have been generated in that range. 

For W0047+68, 15 of the 19 comparison spectra produced outlying pixel shifts. For the comparison spectra that produced outlying pixel shifts, it took an average of 1155 iterations to produce 1000 acceptable pixel shifts; the maximum was 2730 and the minimum was 1001. 

For 2M0241$-$03, six of the 19 comparison objects produced outlying pixel shifts. For two comparison objects, SCH J1612$-$20 and DENIS-P J1605$-$24, restricting the acceptable pixel shifts to ranges of $\pm$50 or fewer re-sampled wavelength pixels still failed to produce a Gaussian distribution of measured RV-induced pixel shifts, so we omitted these two comparison objects in the calculation of the final RV for 2M0241$-$03. For the other four comparison objects, it took an average of 1087 iterations to produce 1000 acceptable pixel shifts; the maximum number of iterations was 1206 and the minimum was 1008. 

For 2M0534$-$06, all of the comparison spectra produced outlying pixel shifts, but with restricting the allowed pixel shift produced Gaussian results for all comparison spectra. It took an average of 6236 iterations to produce 1000 acceptable pixel shifts; the maximum was 59794 and the minimum was 1015. 

\subsection{Space Positions and Motions}
\label{sec:spacemotions}

As yet, no single photometric or spectroscopic youth indicator can be used to assign a precise and reliable age estimate for low mass stars and brown dwarfs. Thus, we rely on kinematics -- positions in RA and DEC (hereafter $\alpha$, $\delta$, and $\pi$); motions in $\mu_{R.A.\cos decl}$, $\mu_{decl}$ (hereafter $\mu_{\alpha*}$, $\mu_{\delta}$ and RV) -- to determine if the brown dwarfs are likely members of a nearby young moving group (NYMG). Even so, kinematics are necessary but not sufficient to prove youth. The sheer number of disk stars and the large kinematic space occupied by these unbound groups mean field-age disk stars can be interlopers, hence the importance of spectroscopic indications that the brown dwarfs are in fact young.

\begin{figure*}
\centering
\includegraphics[width=\textwidth]{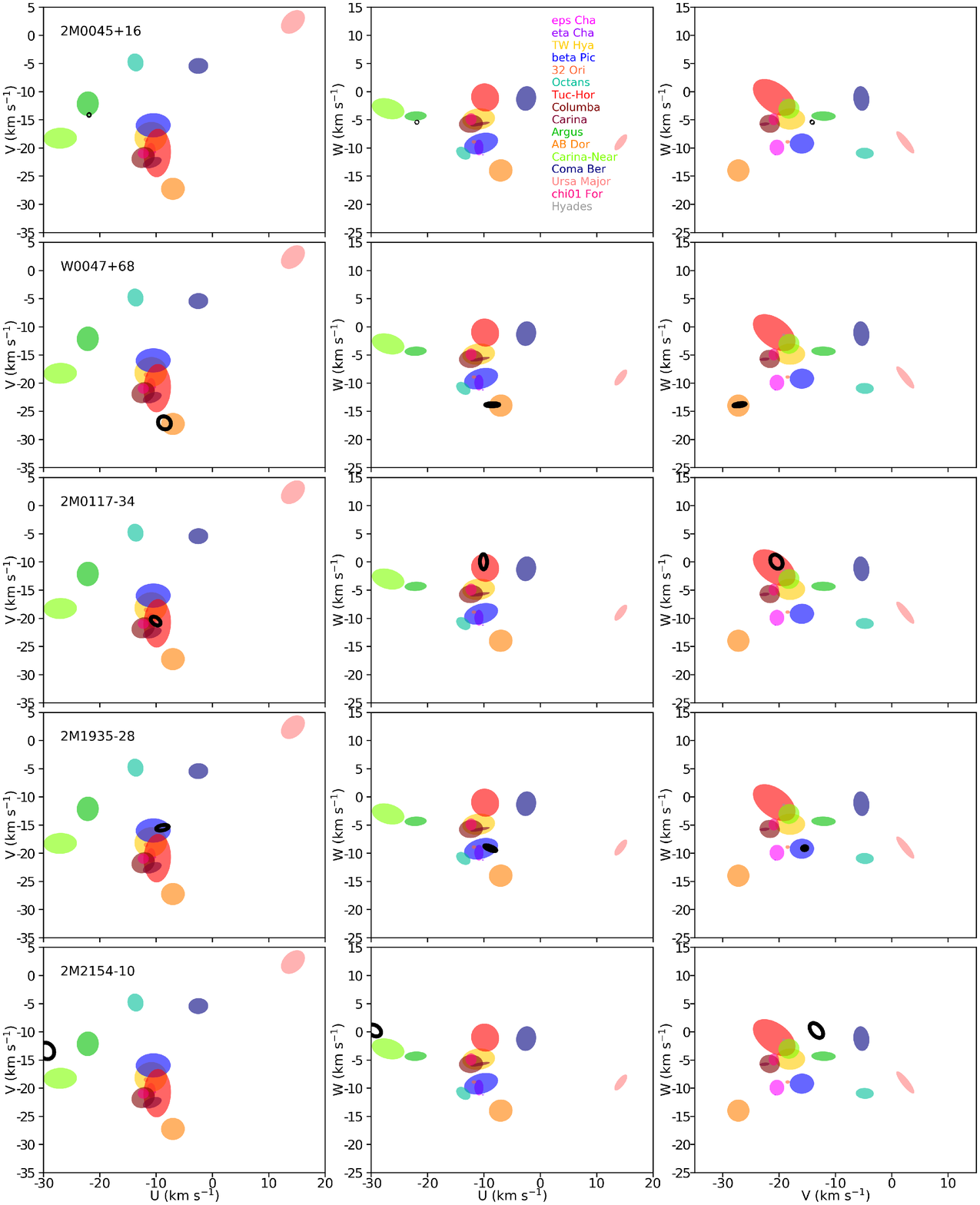}
\centering
\caption{Projected UVW space motions in the UV, UW, and VW planes for the objects from our sample with apparent memberships in a known NYMG. The black ellipse denotes the UVW phase-space position of the object relative to the known NYMGs and nearby open clusters (taken from \citealt{Riedel17}), which are shown with 1-$\sigma$ extents in different colors.}
\label{fig:UVWs_group}
\end{figure*}

\begin{figure*}
\centering
\includegraphics[width=\textwidth]{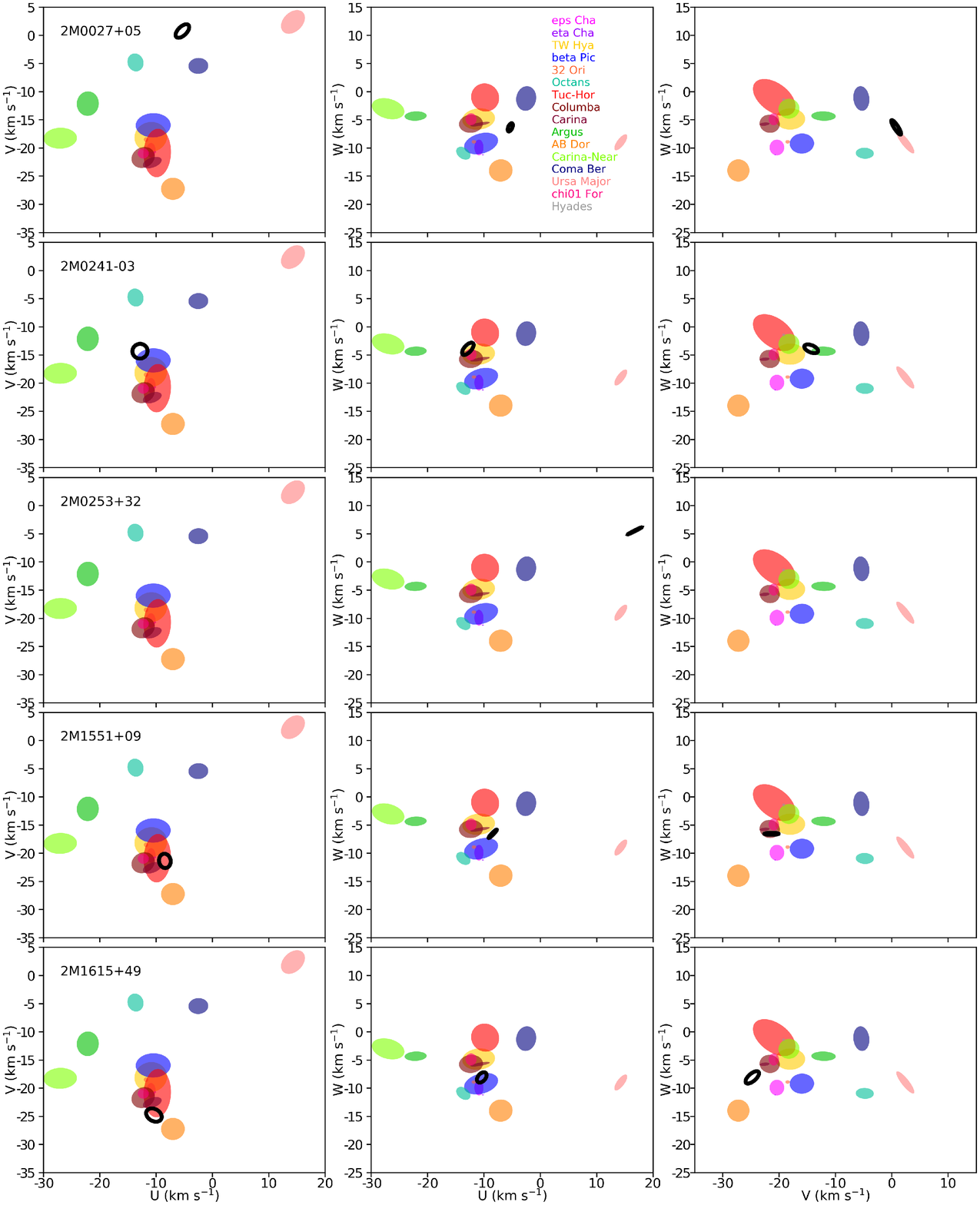}
\centering
\caption{Same as Figure \ref{fig:UVWs_group}, showing the five objects from our sample with parallaxes but no apparent membership in a known NYMG. The V velocity of 2M0253+32 is $-$44.17~$km~s^{-1}$, which is outside the range of our plots.}
\label{fig:UVWs_nogroup}
\end{figure*}

There are two basic strategies for determining membership in moving groups. One is to take the three positional observables ($\alpha$, $\delta$, and $\pi$) and three velocity observables ($\mu_{\alpha*}$, $\mu_{\delta}$, RV), convert them into three-dimensional cartesian space positions (XYZ, where X is toward Galactic center, Y is in the direction of solar motion, and Z is toward the North Galactic Pole) and three-dimensional cartesian space velocities (UVW, where U is motion along the X axis, V along Y, and W along the Z axis), and compare the star's UVWXYZ values to those of the moving group, represented as ellipsoids from \citealt{Riedel17} in Figures \ref{fig:UVWs_group} and \ref{fig:UVWs_nogroup}. The other method is to reverse the process: take the UVWXYZ properties of the moving group, translate them to observable quantities like $\mu$, RV, and $\pi$, at the $\alpha$ and $\delta$ of the target star, and compare the predicted values of a group member to the actual values of the target star. 

It is possible to handle incomplete kinematic data. Indeed, one object in this study lacks a parallax measurement. If converting to UVWXYZ, a range of reasonable parallax values can be tested to see if any are consistent with moving group membership. If converting to observables, it is possible to simply not run a comparison against the predicted parallax value. This ensures that we can still evaluate kinematic memberships, though at the cost of reduced membership certainty. For a more complete discussion of the dependence of membership probabilities on observed data, see \citet{Riedel17}

Positions for our targets come from the 2MASS \citep{Cutri03} catalog, with the sole exception of W0047+68 from WISE \citep{Cutri12}. Proper motions were sourced from a variety of papers, principally \citet{Gagne14a}, \citet{Faherty16}, \citet{Gaia18},  \citet{Casewell08}, and other papers listed in Table~\ref{tab:kinematicdata}. All of our targets have more than one proper motion measurement. Most of the proper motions are relative measurements from catalog surveys or parallax programs and are consistent at the 1-$\sigma$ level for a given object, though all are generally in agreement to within $\pm$10 mas yr$^{-1}$, with the exception of 2M0241$-$03, one uncertain measurement of 2M0117$-$34, and one extremely uncertain measurement of 2M1615+49. We list them all individually in Table \ref{tab:kinematicdata}.

We use published parallax measurements from \citet{Faherty16}, \citet{Dieterich14}, \citet{ZapOso14}, \citet{Dahn02}, \citet{Liu16}, \citet{Gizis15}, and \citet{Gaia18} to obtain complete space motions (Table \ref{tab:kinematicdata}, Figures \ref{fig:UVWs_group} and \ref{fig:UVWs_nogroup}) and more confident group membership probabilities. Seven objects have multiple parallax measurements, which are often discrepant from each other by more than 1-$\sigma$. In the case of 2M0241$-$03, the three parallaxes are only consistent at the 2-$\sigma$ level, and in the case of 2M0045+16, the \citet{Gaia18} parallax is consistent with the \citet{Liu16} parallax but not the \citet{ZapOso14} parallax. In the case of 2M0253+32, the \citet{Faherty16} parallax implies a larger distance than \citet{Liu16} or \citet{Gaia18}, though neither distance makes the target a more likely member of any known NYMG. 2M1935$-$28 is the only case where the \citet{Gaia18} and \citet{Liu16} parallaxes do not agree even at the 2-$\sigma$ level, though both distances independently make the brown dwarf a $\beta$ Pic member. With 2M0027+05, the \citet{Liu16} parallax disagrees with \citet{Dahn02}, though neither parallax produces a likelihood of membership in any known NYMG.

Radial velocities have already been published for 2M0045+16 \citep{Blake10,Faherty16}, W0047+68 \citep{Gizis15}, 2M0241$-$03 \citep{Faherty16}, 2M1615+49 \citep{Faherty16}, and 2M1935$-$28 \citep{Shkolnik17}. Our result for 2M0045+16 is consistent with those measured by \citet{Blake10} and \citet{Faherty16} well within the 1-$\sigma$ uncertainties, even of the most precise measurement ($\pm$0.17~km~s$^{-1}$). Our result for W0047+68 is consistent with the previous measurement to within $\sim$1-$\sigma$ of our lower-precision measurement.
Both of the \citet{Faherty16} measurements were relatively low precision ($\pm$3--8~km~s$^{-1}$), upon which our measurements improve by about a factor of two or more, and our results are consistent to within 1-$\sigma$ uncertainties. 

All of the assembled measurements were combined with standard weighted means and weighted standard deviations. All individual results, and the weighted values (shown in bold) actually used in membership probability analysis, are shown in Table~\ref{tab:kinematicdata}. 


\subsection{Membership Probabilities}
\label{sec:NYMGs}
There are a number of differing approaches to kinematic moving group identification, and following \citet{Faherty16}, we consider the results of five different codes (the four used in \citealt{Faherty16} for comparison purposes, plus the newer BANYAN $\Sigma$ code) to provide consensus approach. Four of the codes used here (BANYAN I, BANYAN II, BANYAN $\Sigma$, and LACEwING, see descriptions and references below), test against different properties: proper motion, radial velocity, parallax, and space position. The fifth code, the convergence code from \citet{Rodriguez13}, considers only a single test for proper motion, but predicts the distance and radial velocity. If data do not exist or are not present, the test is simply not run and the codes produce results based only on the other tests. 

The Convergence code presented in \citet{Rodriguez13} exploits the fact that if all of the stars in a moving group exhibit parallel space motions, their proper motion vectors should converge at a point in space (a "convergent point", analogous to the vanishing point) because of perspective effects. The code computes probabilities of membership in six moving groups (TW~Hya, $\beta$~ Pic, Tuc-Hor, Columba, Carina-Near, and AB~Dor) by comparing the proper motion vector defined by ($\mu_{\alpha*}$, $\mu_{\delta}$) to one pointing toward the convergent point of a given moving group. From there it predicts the associated radial velocity and distance of the object, which can be compared to any actual measurements.

BANYAN~I \citep{Malo13} uses a Bayesian formalism to evaluate which of seven nearby young moving groups (TW~Hya, $\beta$~Pic, Tuc-Hor, Columba, Carina, Argus, and AB~Dor) or a field population of which an object is most likely to be a member. It converts observables to Cartesian space. Unlike the Convergence code, radial velocity and parallax measurements are incorporated into the probability distribution rather than predicted according to possible group membership. 

BANYAN~II \citep{Gagne14a} is a modification of BANYAN~I. It considers the same seven moving groups as BANYAN~I, but it is based on a different set of bona-fide members, assumes an uneven distribution of the number of stars in each group, and allows freely oriented moving groups in space. It takes a hybrid approach, constraining observables based on Cartesian space.

LACEwING \citep{Riedel17} predicts memberships in 16 nearby young moving groups and open clusters within 100 pc: compared to the BANYAN codes, it adds $\epsilon$~Cham\ae leon, $\eta$~Cham\ae leon, 32~Orionis, Octans, Carina-Near, Coma Berenices, Ursa Major, $\chi^{01}$~Fornax, and the Hyades. Like BANYAN~II, all groups are represented as freely-oriented ellipsoids with numerically proportionate populations. Unlike BANYAN~II, it does not use Bayesian priors; instead it relies on the parameterized results of a large simulation of stars to translate goodness-of-fit values into membership probabilities. It operates in observational space.

BANYAN~$\Sigma$ \citep{Gagne18} is a more refined version of BANYAN~II using multivariate Gaussian models (instead of the orthogonal axis models of BANYAN II and LACEwING) which predicts memberships in 27 young moving groups and associations with ages up to 800~Myr and distances up to 150~pc, including all of the groups in LACEwING, plus 118~Tau, Corona Australis, Upper Corona Australis, IC~2391, IC~2602, Lower Centaurus Crux, Upper Centaurus Lupus, Upper Sco, $\rho$~Oph, the Pleiades, Taurus, Platais~8, Volans-Carina, and the new formulation of Argus identified in \citet{Zuckerman19}. Like BANYAN~II, it constrains the observables based on Cartesian space.

\section{Membership Results and Notes on Individual Objects}
\label{sec:notes}
We identify five of our 11 sample objects as high probability members of known NYMGs. Four of these are re-confirmations of possible memberships presented in \citet{Gizis15}, \citet{Faherty16}, \citet{Liu16}, or \citet{Shkolnik17}, and one is a new membership. The remaining six objects are found to have no membership in a known NYMG.

\paragraph{2M0045+16} (L2$\beta$) was identified by \citet{Gagne14c} as a member of the roughly 50 Myr old Argus association, using more or less the original definition of Argus from \citet{Torres08} and by \citet{Liu16} using BANYAN~II. That identification is reconfirmed here with 90-100\% probabilities, which maintains this object as one of the few brown dwarfs in Argus, with an estimated mass of 25.0$\pm$4.6 M$_{jup}$ \citep{Faherty16}. Given that Argus has been kinematically \citep{Torres08} and chemically \citep{deSilva13} associated with the nearby IC 2391 open cluster, we can draw on the properties of hundreds of higher mass stars to understand this and other similarly young brown dwarfs. 2M0045+16 is a member of Argus, both in its original formulation (\citealt{Torres08} and subsequent, used in BANYAN I, BANYAN II, and LACEwING) which was disputed by \citet{Bell15}, and the new definition from \citet{Zuckerman19} (used in BANYAN $\Sigma$).

There are three published parallaxes for 2M0045+16, two of which agree with each other, while a third value from \citet{ZapOso14} is inconsistent at a 2-$\sigma$ level (see Table~\ref{tab:kinematicdata}). Even using discrepant parallax, we find Argus to be the most likely NYMG membership by all methods that consider membership in Argus.

As alternative hypothesis, LACEwING suggests that 2M0045+16's kinematics are also consistent with $\beta$~Pictoris. This would make 2M0045+16 a significantly younger brown dwarf of roughly 25~Myr \citep{Mamajek14} rather than 50~Myr.

We propose the 2MASS~J0045+16 be considered an RV standard given the precision and stability of measurements from three different studies and its relative brightness among early L dwarfs. Its $J$=13.06 mag makes it the 6th brightest L2 and in the top 25 brightest early ($<$L5) L dwarfs, just 0.6~mag fainter than the brightest known L2 and $\sim$1~mag fainter than the brightest single early L~dwarf. RV measurements have previously been reported by \citet{Blake10} from 2003 $K$-band observations (3.29$\pm$0.17~km~s$^{-1}$) and \citet{Faherty16} from 2008 $H$-band observations (3.16$\pm$0.83~km~s$^{-1}$), both also from NIRSPEC. Given the consistency of these and our measurement of 3.29$\pm$1.33 from 2014 NIRSPEC $J$-band observations, it seems that 2MASS~J0045+16 is RV stable and an optimal late-type spectral standard.

\paragraph{W0047+68} (L7$\gamma$) has previously been identified as an AB~Doradus member by \citet{Gizis15} and \citet{Liu16} using full UVWXYZ space motion and position fitting. We reconfirm that membership: W0047+68 is a member of AB~Doradus according to every code, despite a 2$\sigma$ disagreement between our radial velocity and that of \citet{Gizis15}. This L7$\gamma$ object is one of the least massive known free-floating extrasolar objects, with an estimated mass of 11.8$\pm$2.6 M$_{Jup}$ \citep{Faherty16}, despite being substantially older than other brown dwarfs with a $\gamma$ gravity classification.

\paragraph{2M0117$-$34} (L1$\beta$) is confirmed with our RV measurement and the parallax from \citet{Gaia18} as a member of Tucana-Horologium, an identification made by \citet{Faherty16} solely on the basis of its proper motion and by \citet{Liu16} on the basis of its proper motion and parallax. The new membership is agreed upon by every moving group code and implies that the brown dwarf is 16.4$\pm$3.7 M$_{Jup}$ \citep{Faherty16}. The alternative proper motions from \citet{Casewell08}, \citet{Gagne14a} and \citet{Liu16} have much larger motion along the $\alpha$ axis than the proper motion calculated by \citet{Faherty16}, but all the membership codes still find membership in Tucana-Horologium.

\paragraph{2M1935$-$28} (M9$\gamma$) is a member of $\beta$~Pictoris, first identified as such by \citet{Shkolnik17}. With LACEwING, BANYAN~I, and BANYAN~$\Sigma$, it is a moderate or high probability member; with BANYAN~II, it is lower likelihood; and with the Convergence method, it is either a $\beta$~Pictoris or Columba member (see Table \ref{tab:memberships}). We consider this system as a high probability member of $\beta$~Pictoris.

\paragraph{2M2154$-$10} (L4$\beta$) is identified by LACEwING, the Convergence Code, and BANYAN $\Sigma$ as moderate probability member of Carina-Near, a 200 Myr old group identified by \citet{Zuckerman06}. This makes 2M2154$-$10 the oldest confirmed NYMG member in the sample.
\citet{Gagne14a} found 2M2154$-$10 to be a member of Argus, which we do not reproduce due to a disagreement in the RV: as a member of Argus it should have an RV of roughly $-$14 km s$^{-1}$, while we measure an RV of $-$21$\pm$2~km s$^{-1}$.

\paragraph{The remaining targets} (2M0027+05 [M9.5$\beta$], 2M0253+32 [M7$\beta$], 2M0534$-$06 [M8$\gamma$], 2M0241$-$03 [L0$\gamma$], 2M1551+09 [L4$\gamma$], and 2M1615+49 [L4$\gamma$] were all identified having ambiguous NYMG membership by \citet{Faherty16} and are not found to be likely members of any known NYMG with the addition of our RV measurements and \citet{Gaia18} astrometry. 

The convergence method predicts that 2M0253+32 is a member of $\beta$~Pic with a predicted RV of 5.5~km~s$^{-1}$, but our measured RV is $-$34.5~km~s$^{-1}$. We therefore conclude that 2M0253+32 is not a $\beta$~Pic member.

2M1615+49 appears to be a rapid rotator (see Figure~\ref{fig:SpT_FWHM}) but does not otherwise distinguish itself. LACEwING finds it to be a potential member of AB Doradus, though at low probability; the Convergence code finds it a possible member of Tuc-Hor (with a predicted RV of $-$15~km~s$^{-1}$, which does not match our measured $-$24~km~s$^{-1}$), and BANYAN I, BANYAN II, and BANYAN $\Sigma$ find it is not a member of any group at a probability above the threshold of interest.

The expected RV for 2M1551+09 if it were a member of $\beta$~Pic, $-$17~km~s$^{-1}$, is consistent with the actual measured velocity of $-$15~km~s$^{-1}$, but only the convergence code finds that membership and at a low probability. The expected distance for a $\beta$~Pic member with the proper motion of 2M1551+09 would place it very far spatially from the known members of $\beta$~Pic (a condition the BANYAN codes and LACEwING consider), which means that even if its parallax-determined distance matches the expected distance of 30~pc, the object cannot be a member. 

2M0241$-$03 has five published proper motions \citep{Faherty16,ZapOso14,Gagne14a,Casewell08,Liu16} and three parallaxes \citep{Faherty16,ZapOso14,Liu16}, which only agree with each other at the 2$\sigma$ level. This system has been considered a member of Tucana-Horologium since \citet{Gagne14a}, but with our weighted parallax we find no such membership. Using the \citet{Faherty16} and \citet{ZapOso14} parallaxes individually, the brown dwarf is still not a member of any moving group. \citet{Liu16} placed it in Tuc-Hor with an 82\% likelihood using BANYAN~II (a lower probability than our threshold for BANYAN~II) using parallax and proper motion, but with our astrometry BANYAN~II gives us an 88\% membership (also below the threshold) in $\beta$ Pic instead. LACEwING does reproduce membership in Tuc-Hor at a low 30\% probability, and we note that LACEwING gives a higher (46\%) chance of membership in Columba. The Convergence code suggests a low probability of membership in Carina-Near if the system is at 80 pc, which it is not.

Ultimately, the reason it is not in Tuc-Hor is a combination of factors: if its (combined) proper motion and radial velocity were to imply the best possible space velocity match to Tuc-Hor, the brown dwarf would need to be closer to 64 pc away, which not even the \citet{Liu16} parallax (54 pc) agrees with, while (simultaneously) being at that appropriate distance would put it approximately 40 pc away (over 2-$\sigma$) from the bulk of the Tuc-Hor moving group.

For now, we suspect that these objects are members of a young field population, which \citet{Riedel17} has shown to be quite substantial.

\begin{figure}
\centering
\includegraphics[width=0.5\textwidth]{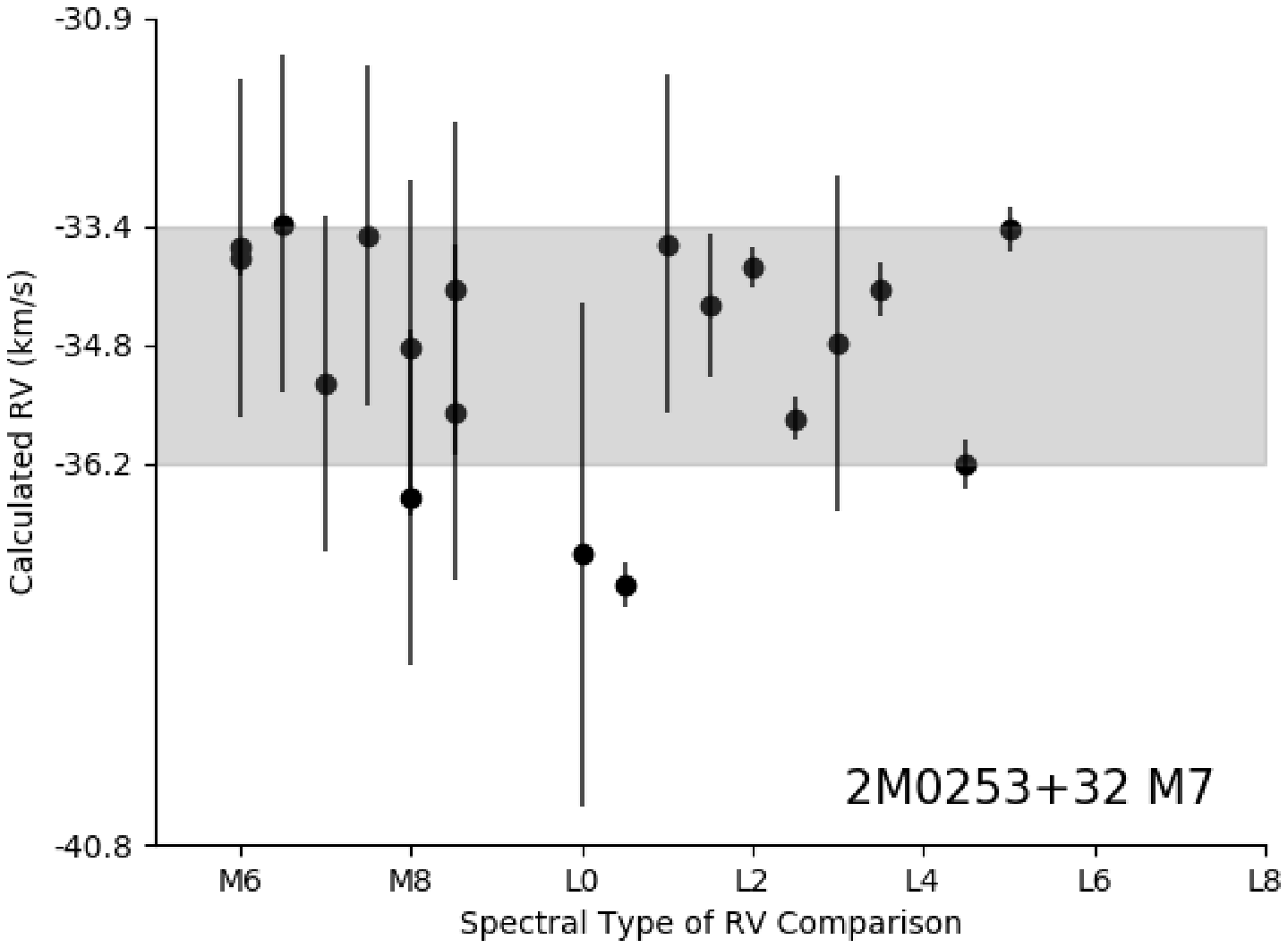}
\includegraphics[width=0.5\textwidth]{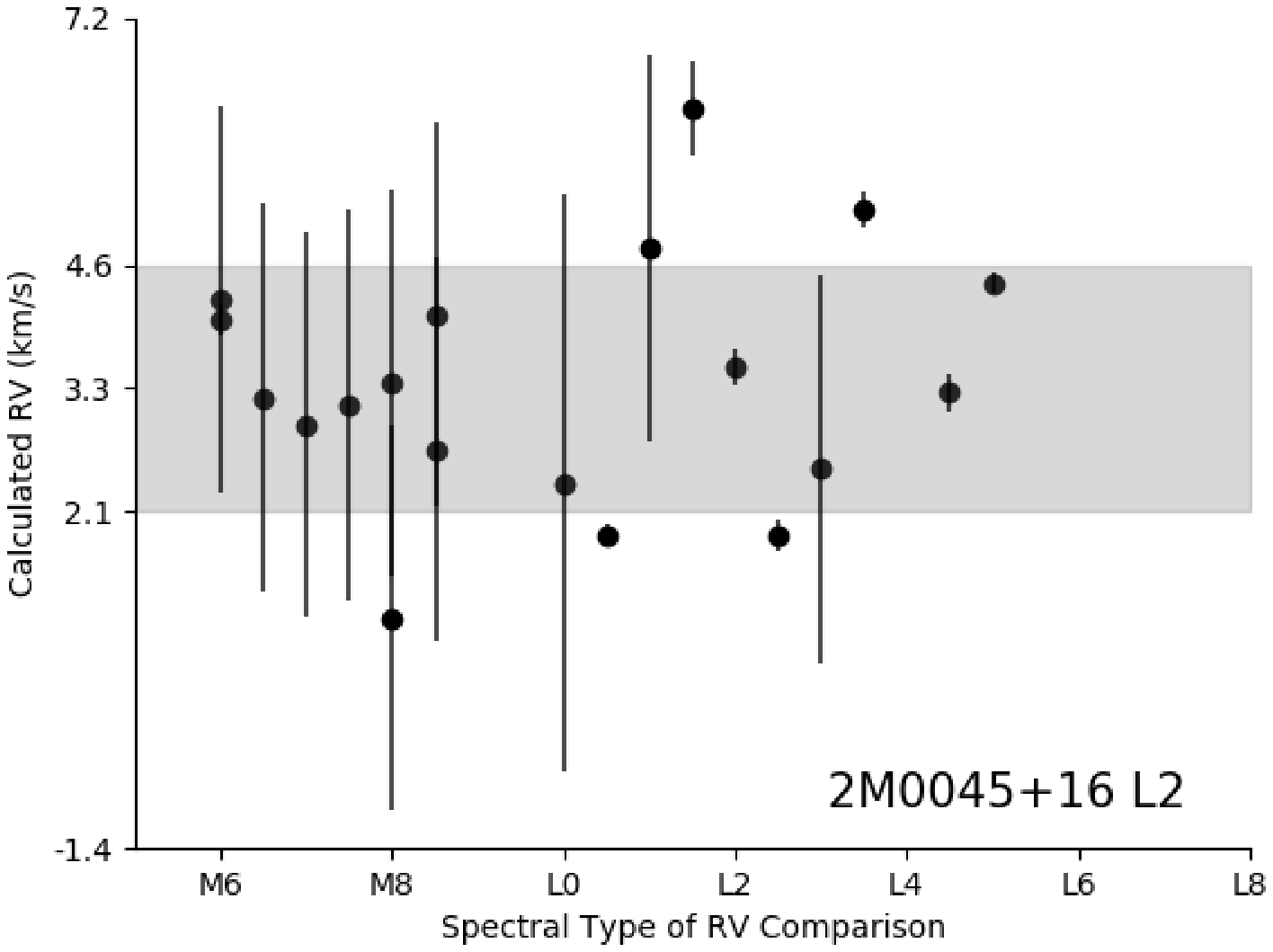}
\includegraphics[width=0.5\textwidth]{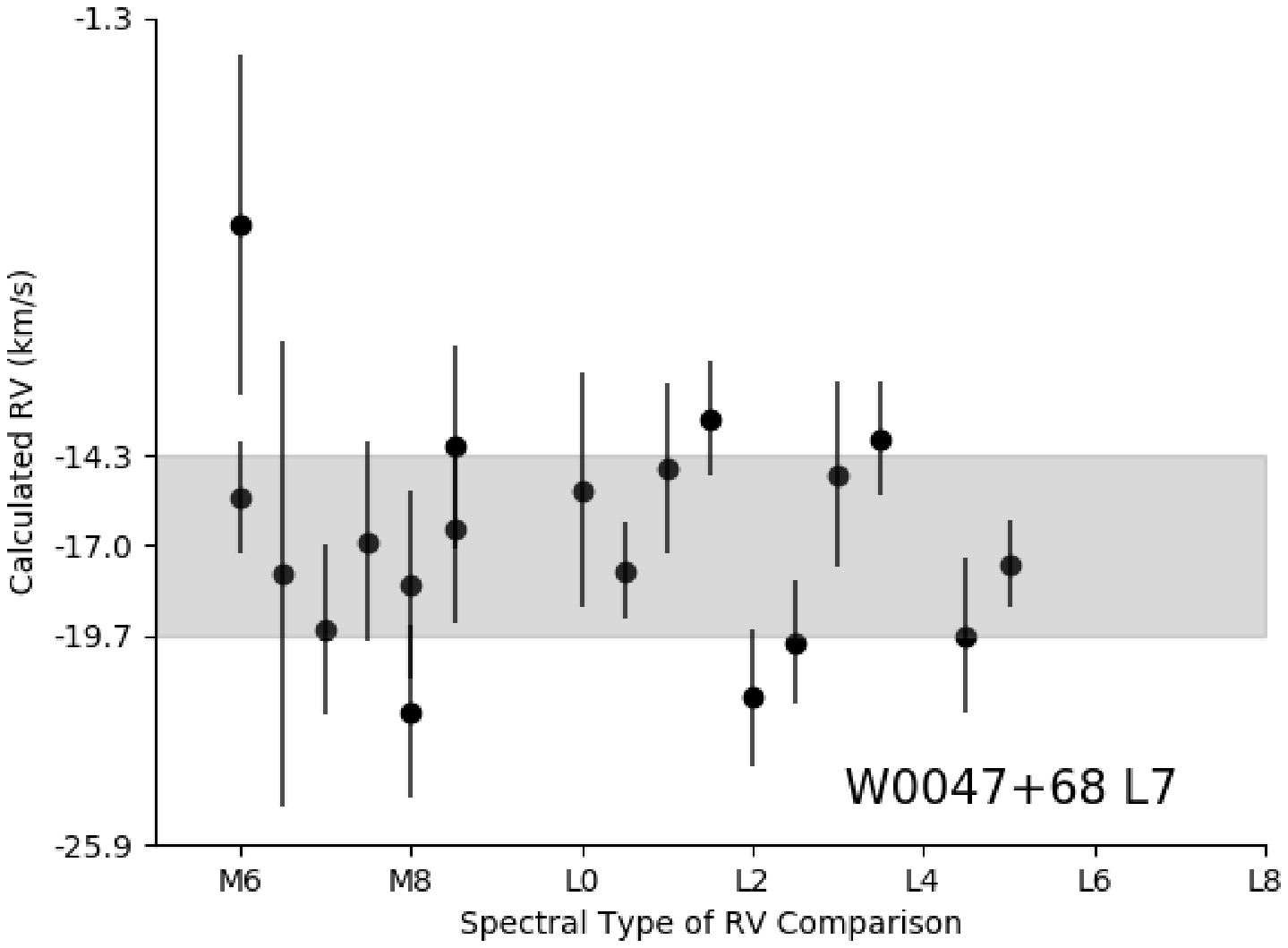}
\centering
\caption{Measured RV (km~s$^{-1}$) versus spectral type of each comparison object for 2M0253+32 (top), 2M0045+16 (middle), and W0047+68 (bottom). The gray bar represents the RV 1-$\sigma$~uncertainty range on the final results for each target. The RV of the target is calculated by cross-correlating each target and comparison pair, and does not show any correlation to or dependence on the difference in spectral type between the comparison and target objects.}
\label{fig:UncSpT}
\end{figure}

\section{\sc Discussion}
\label{sec:discussion}

\subsection{RV Measurements of Very-Low-Mass Objects}
\label{sec:RVtests}

Typically, high-resolution spectra are cross-correlated against spectra of objects with similar spectral types, effectively doubling the required observing time, which can be on the order of several hours for intrinsically faint very-low-mass stars and brown dwarfs. For example, \cite{Prato15} used a similar cross-correlation method to ours for measuring the radial velocities of very-low-mass objects, but restricted that comparison to objects with similar spectral types. In order to optimize the efficiency of RV measurements for very-low-mass objects, we test the dependence of the precision and accuracy of RV results on the spectral type of the comparison object.

Figure~\ref{fig:UncSpT} shows the spectral type of each RV comparison, as listed in Table~\ref{tab:standards}, as a function of the calculated RV of the target object. The gray bar indicates the 1-$\sigma$~uncertainty on the final RV measurement of the target, as listed in Table \ref{tab:kinematicdata}. The three panels show the objects of earliest and latest spectral type in our sample, 2M0253+32 (M7, top panel) and W0047+68 (L7, bottom panel), as well as the object of median spectral type, 2M0045+16 (L2, middle panel). The same test was done for all eleven objects in our sample. We see no correlation between spectral type of the comparison object and the precision or accuracy of the individual RV measurement, compared to the final value. Thus, we show that a cross-correlation comparison objects can be as different as $\pm$5 spectral types from the target object. This is similar to the findings of \citet{Newton14} for M dwarfs (their Section~8.2).

These results can improve the efficiency of observations required for using the cross-correlation technique for measuring the RVs of late-type objects by loosening the requirements for comparison object's spectral type similarity to that of the target. If a close spectral type match is not required, less observing time needs to be spent on assembling (from previously-obtained spectra, or new observations) a library of high-resolution comparison spectra, and the expense of making RV measurements of late-type objects decreases markedly.

\subsection{NIR Colors of Young Very-Low-Mass Objects}
\label{sec:nir}
By virtue of their selection as objects with spectral indicators of youth, most of our sample has redder near- and mid-infrared colors than expected for normal L dwarfs (Figure~\ref{fig:JW1}), with the exception of 2M0253+32, which is consistent with the colors of a normal L dwarf of the same spectral type. The degree of reddening is fairly consistent across all near- and mid-infrared color combinations, with a few exceptions:
In $H-K$, 2M0027+05 is bluer than (but consistent with) a normal L dwarf; in $K-W1$, 2M0534-06 is likewise bluer or consistent with a normal L dwarf. Neither of these effects are due to poor precision photometry; they appear to be real (or perhaps variable) features of the objects themselves.

The NIR colors alone are not a sufficient gauge of age. The most consistently discrepant objects in the sample, which are also generally the objects with the largest color offsets from normal brown dwarfs, are 2M0241$-$03, 2M1615+49, 2M2154$-$10, and W0047+68; W0047+68 is identified as an AB~Dor (120~Myr) member and is both the potentially oldest identified member in the sample and the coldest brown dwarf in the sample, while the newly-identified $\beta$ Pictoris (25 Myr) member, 2M1935$-$28, is just above the envelope of young brown dwarfs in Figure \ref{fig:JW1}. 
\begin{figure}
\centering
\includegraphics[width=0.5\textwidth]{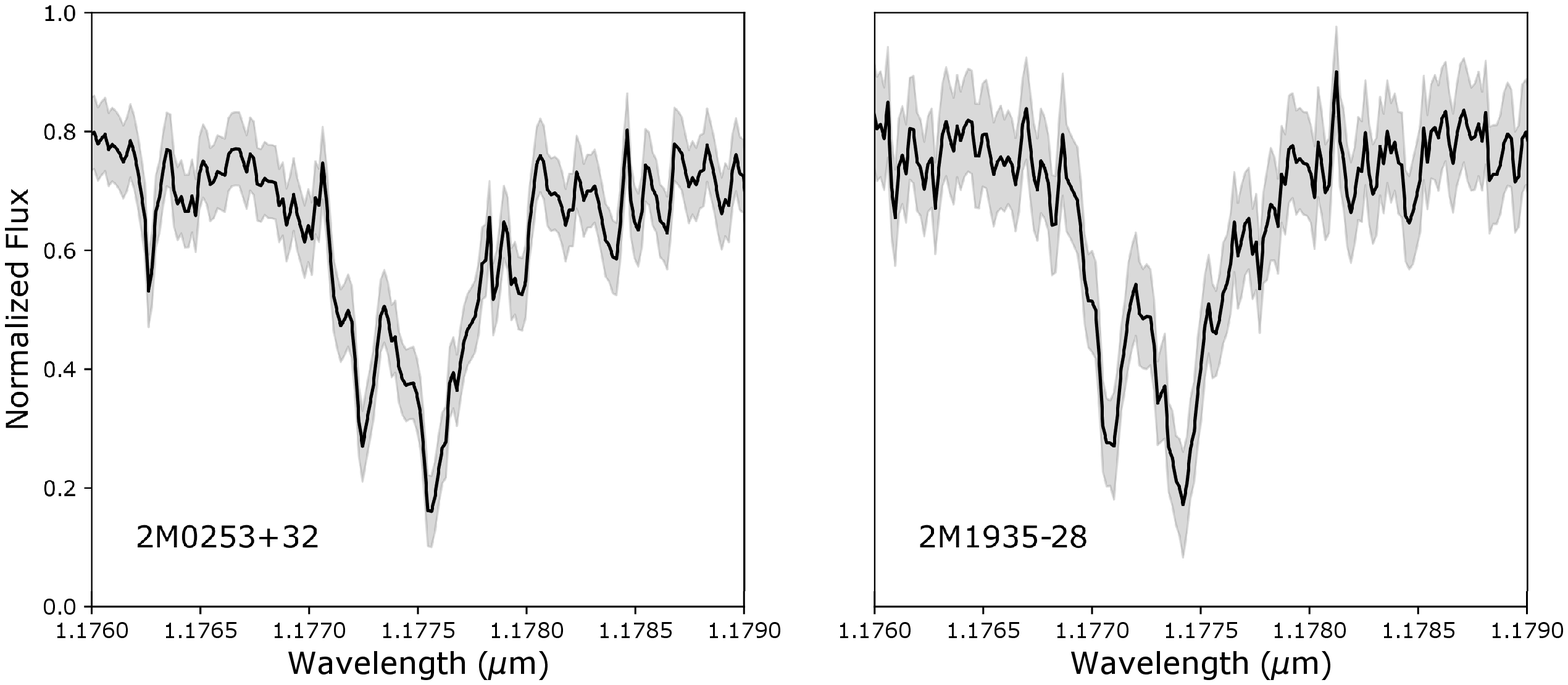}
\centering
\caption{K~{\sc i} triplet for 2M0253+32 (left panel) and 2M1935$-$28 (right panel).}
\label{fig:KI_Triplet}
\end{figure}

\subsection{K~{\sc i} Line Strengths}

Measurement of gravity-sensitive spectral lines may provide a more reliable indicator of youth than NIR colors alone \citep[e.g.,][]{Faherty12}. The neutral alkali metal absorption lines like those of Na~{\sc i} and K~{\sc i} are weaker in lower surface gravity atmospheres \citep[e.g.,][]{Schlieder12}, which translates to smaller equivalent widths (EWs). This provides a way to test if red objects are truly young and low surface gravity, or simply red because of dustier atmospheres. With our high-resolution spectra we can measure the strength and width of gravity-sensitive lines and test this directly, which we describe below. We can also determine if these objects are rapid rotators by measuring the Full Width at Half Maximum (FWHM) of the lines.

NIRSPEC orders 61 (1.24--1.26~$\mu$m) and 65 (1.165--1.182~$\mu$m) each contain a K~{\sc i} doublet (in order 65, the lines are sometimes resolvable into a triplet, see Figure~\ref{fig:KI_Triplet}) that has been shown to be sensitive to temperature \citep[e.g.,][]{McLean07} and surface gravity \citep[e.g.][]{McGovern04,Rice10}. Therefore we measure these line strengths and compare them to those of field-age objects to evaluate additional indicators of youth for our sample. Because of the only occasionally resolved triplet in order 65, which is also typically of lower SNR, we concentrate our analysis on the order 61 doublet.

\begin{figure}
\centering
\includegraphics[width=3.7in]{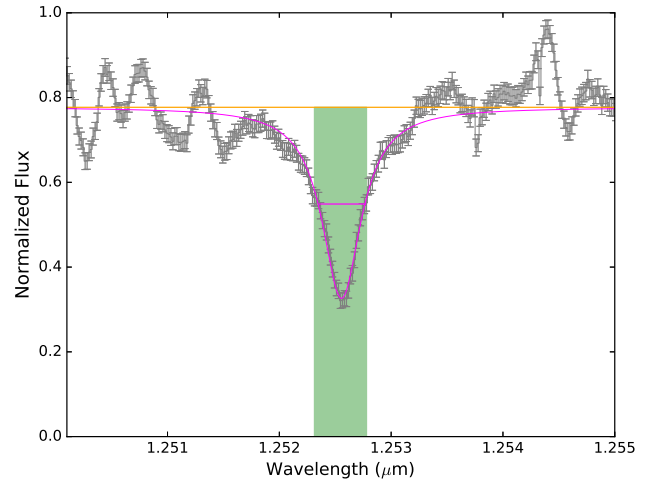}
\centering
\caption{Subsection of NIRSPEC dispersion order 61 for 2M0045+16 centered on the $\lambda$=1.2525 K~{\sc i} line to demonstrate the line strength measurement methods of \citet{Alam16}. The yellow line represents the defined pseudocontinuum, the purple curve the Voigt profile, and the purple horizontal line the FWHM. The shaded green region represents the equivalent width.}
\label{fig:PHEW2m0045_61b}
\end{figure}

\begin{figure}
\centering
\includegraphics[width=3.7in]{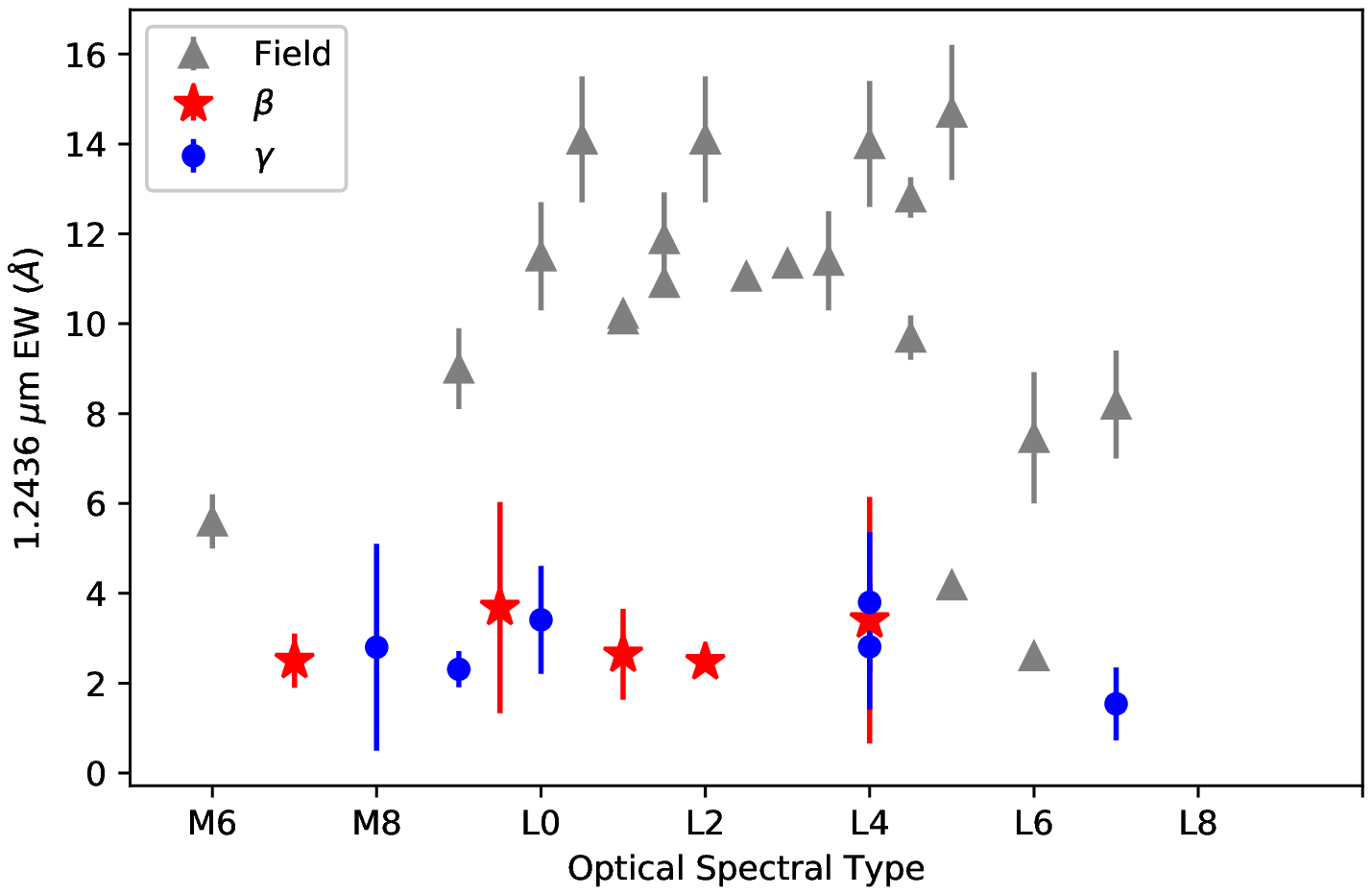}
\includegraphics[width=3.7in]{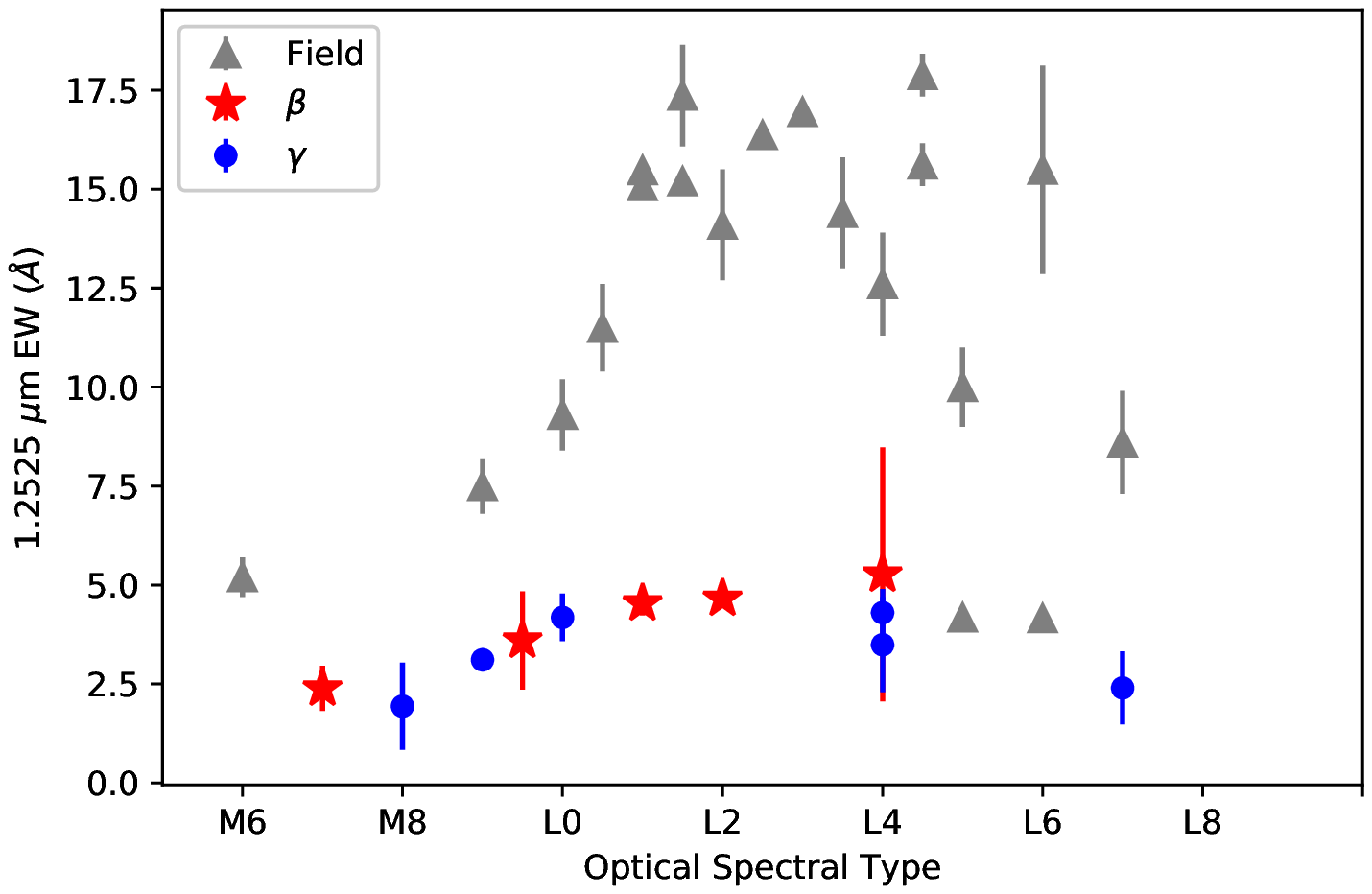}
\centering
\caption{K {\sc i} equivalent width versus optical spectral type for our sample of candidate young, unusually red objects organized by gravity class (red stars, blue circles) compared to field objects (gray triangles) for the order 61 lines at 1.2436 $\mu$m (top panel) and 1.2525 $\mu$m (bottom panel) lines.}
\label{fig:SpT_EW}
\end{figure}

\begin{figure}
\centering
\includegraphics[width=3.7in]{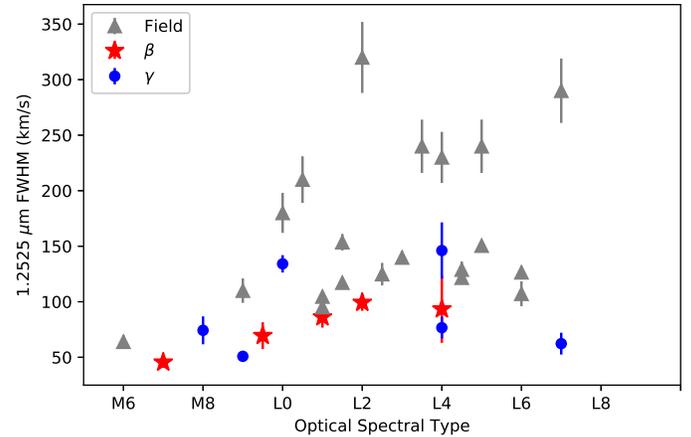}
\centering
\caption{K {\sc i} FWHM versus optical spectral type for our sample of candidate young, unusually red objects compared to field objects for the order 61 line at 1.2525 $\mu$m line. Colors are the same as Figure \ref{fig:SpT_EW}.}
\label{fig:SpT_FWHM}
\end{figure}

Following the methods of \citet{Alam16}, we quantify the strengths of the $\sim$1.25~$\mu$m K~{\sc i} lines for our sample by computing EWs and FWHM using PHEW: PytHon Equivalent Widths\footnote{\url{https://github.com/munazzaalam/PHEW/}} code. We measure EWs using $0^{th}$-order fit to the pseudo-continuum, defined as the average flux outside of the absorption line within a 1.241--1.246 $\mu$m window, and a Voigt profile fit to the absorption line. The equivalent width is calculated by integrating the pseudo-continuum level minus the spectrum over the selected range. Uncertainties were estimated via 1,000 Monte Carlo iterations. We report the means and standard deviations of these measurements in Table \ref{tab:phew_tab}. Figure~\ref{fig:PHEW2m0045_61b} presents an example of these measurements for the $\lambda$=1.2525 K~{\sc i} line from the spectrum of 2M0045+16. Line strength measurements, compared with results for field objects from Alam et. al (in prep.) and \citet{McLean03}, are presented in Figures \ref{fig:SpT_EW} and \ref{fig:SpT_FWHM}. The complete dataset for both field and suspected young objects is presented in Table \ref{tab:phew_tab}.

These results follow the general pattern indicated by other studies of the K~{\sc i} lines, e.g. \citet{Allers13} (Figure 23), \citet{Gagne15a} (Figure 6), and \citet{Martin17} (Figure 3) and indicate that our suspected young sample exhibits lower surface gravities, as expected for objects with $\beta$ and $\gamma$ gravity designations. Our results are not directly comparable to those of the aforementioned papers due to our higher spectral resolution. Those papers used moderate (R$\sim$750--2000) spectra, while our R$\sim$20,000 spectra yields more precise measurements, higher EWs for field stars, a decreased sensitivity to the FeH feature overlapping with the 1.2436 $\mu$m line and correspondingly more distinction between field objects and our suspected young sample, even at the extremes where \citet{Allers13} could only determine an EW-based gravity classification and saw no differences between young and field stars in their K~{\sc i} index.

The FWHM measurements (Figure \ref{fig:SpT_FWHM}) demonstrate that almost all of our targets have lower $v\sin{i}$ than the field objects. Two objects, 2M0241$-$03 (L0$\gamma$) and 2M1615+49 (L4$\gamma$), are possibly rapid rotators and/or viewed more edge-on than the other young objects, which would broaden their gravity-weakened lines, as evidenced by their higher FWHM measurements but similar EWs to the other candidate young objects.

\subsection{Consistency of Age Indicators}
\label{sec:consistency}

Near-infrared colors may indicate possible youth, but they have never been considered sufficient to determine specific ages for young brown dwarfs, as noted in Section \ref{sec:nir}. Spectroscopic measurements are more useful for evaluating youth, but here too there are limitations. All of our objects have been classified as either $\beta$ or $\gamma$ gravity classes according to their red-optical spectra (Table \ref{tab:observations}), and most objects have INT-G or VL-G gravity classes on the near-infrared \citet{Allers13} spectral system. As shown in Figure \ref{fig:SpT_EW}, all of our targets have weaker (lower EW) K~{\sc i} doublet lines than field-age dwarfs of comparable spectral type, indicating youth, though there is substantial overlap between $\beta$ and $\gamma$ gravity classifications. Furthermore, gravity-related spectral type suffixes themselves do not appear to track directly with age; our ~50 Myr old Argus member 2M0045+16 is an L2$\beta$, while our ~125 Myr old AB Doradus member W0047+68 is classified as L7$\gamma$. We therefore cannot assign even relative ages based on line strength or gravity measurements alone. We also do not consider it problematic that W0047+68 shows signs of youth when M dwarf members of AB Doradus typically do not have identifiable low surface gravity features \citep{Schlieder12}, as it is a much cooler, thus lower-mass, object that may be evolving more slowly. 

Even the non-spectrophotometric property of kinematic memberships has limitations. Taken in total, the line strengths and colors indicate that all the objects in our sample are young (if not precisely how young), but we can only connect five of them with NYMGs that confirm a young age. The failure to connect the remaining objects to a NYMG could be explained by one of four possibilities:
\begin{enumerate}
    \item The NYMG identification algorithm may be flawed because it is based on inaccurate or incomplete assumptions of how to best identify NYMGs.
    \item We may have insufficiently precise kinematic data for the late-type object, or an inaccurate understanding of the parameters of the NYMG itself.
    \item The object may be a member of an as-yet-unknown NYMG.
    \item The object may be a young unassociated or field object, for example the product of a one-off star formation event, as suggested by \citealt{Riedel17} and \citealt{Gagne18}.
\end{enumerate}
We can attempt to explore the first two possibilities by using multiple techniques: the five different moving group identification codes all have different algorithms and different parameterizations of the NYMGs. Even so, the codes agree that most of our targets are not members of any known NYMG, strengthening the probability of the third and/or fourth options.

The current situation is thus: We have photometric color, gravity classification, and individual line strengths (EWs and FWHM). They frequently disagree with each other about the degree to which an object is young, particularly when compared to the age implied by kinematic membership in a NYMG. There is as yet no simple spectrophotometric indicator (or group of indicators) that reliably indicates age, so we are still limited to saying that by our aggregate analysis of the various metrics, the objects are likely young. Only the five objects we can connect to a NYMG can give us an age; for the other six, all we can say is that they are likely young, and there is no reason to suspect they are not as young as the confirmed NYMG members in the sample.

\section{\sc Conclusions}
\label{sec:conclusions}
In this paper we have presented new high-resolution NIR spectroscopy of 11 red, low-gravity, late-type objects. Using new RV measurements derived from that spectroscopy, and proper motion and parallax measurements from literature sources, we re-confirm membership of four objects in NYMGs. We also identify a new member of Carina-Near and confidently rule out six objects as members of the known NYMGs. These objects remain interesting targets for study, though we cannot currently determine their ages or origins.

Our study also adds more evidence to the hypothesis (such as proposed by \citealt{Riedel17} and \citealt{Gagne18}) that there are other populations of young objects in the solar neighborhood yet to be discovered, whether they are new NYMGs or a genuinely unassociated "field" population of young objects. The six objects we conclusively rule out as members of the known NYMGs are an indistinguishable population, spectroscopically and photometrically, from the confirmed NYMG members. There is no reason to say that they are not young, beyond lack of group membership.

We also presented evidence that the accuracy of the cross-correlation technique is not dependent on close spectral type matches. Previously, it was thought that spectral types of standard stars had to be as close as possible to the spectral type of the target - within two subtypes - for the cross-correlation radial velocity technique. The power of this technique in face of the spectral type discrepancy is due to the strength and regularity of the FeH lines in cool star spectra, the rectification step where a third-order polynomial fit corresponding to the overall shape of the spectrum is removed, and the use of multiple comparison spectra in a weighted measurement. The end result is proof that collecting an extensive library of standards at every spectral type is not necessary to achieve kilometer-per-second precision radial velocities, and therefore shows that the technique of cross-correlation is cheaper and easier to implement than previously thought.

\acknowledgments

The authors wish to thank the staff of the Keck Observatory for their outstanding support, including Luca Rizzi, Jim Lyke, Cynthia Wilburn, Terry Stickel, Jason McIlroy, Heather Hershley, and Barbara Schaefer. Observing assistance from Kay Hiranaka was greatly appreciated. The authors are grateful for assistance with references from J.T. Wright and E.R. Newton via Twitter, and J. Gagne for help with the BANYAN II and BANYAN $\Sigma$ codes. ARR, VD, and ELR were responsible for writing the majority of the paper. ARR was responsible for the youth analysis, conclusions, and system notes. VD and EA were responsible for the NIRSPEC data reduction and description in the paper. VD, ELR, EA, and ARR were responsible for the RV analysis. MKA was responsible for the line measurements, and description of PHEW. KLC was responsible for the spectral typing and advising of VD and EA. JKF was responsible for the general survey outline. 

This material is based upon work supported by the National Science Foundation under grant numbers AST-1313278, AST-1313132, and AST-1153335. EA acknowledges support from the National Science Foundation Graduate Research Fellowship under Grant No. DGE 1752814. This research was supported in part by NASA through the American Astronomical Society's Small Research Grant Program. This research has made use of the NASA/IPAC Infrared Science Archive, which is operated by the Jet Propulsion Laboratory, California Institute of Technology, under contract with the National Aeronautics and Space Administration. This publication makes use of data from the Two Micron All Sky Survey, which is a joint project of the University of Massachusetts and the Infrared Processing and Analysis Center, funded by the National Aeronautics and Space Administration and the National Science Foundation. This research has benefited from the M, L, and T dwarf compendium housed at DwarfArchives.org and maintained by Chris Gelino, Davy Kirkpatrick, and Adam Burgasser. This research has made use of the SIMBAD database, operated at CDS, Strasbourg, France and NASA's Astrophysics Data System. his work was supported by a NASA Keck PI Data Award, administered by the NASA Exoplanet Science Institute. Keck telescope time was granted by NOAO, through the Telescope System Instrumentation Program (TSIP). TSIP is funded by NSF. Data presented herein were obtained at the W. M. Keck Observatory from telescope time allocated to the National Aeronautics and Space Administration through the agency's scientific partnership with the California Institute of Technology and the University of California. The Observatory was made possible by the generous financial support of the W. M. Keck Foundation. The authors wish to recognize and acknowledge the very significant cultural role and reverence that the summit of Maunakea has always had within the indigenous Hawaiian community. We are most fortunate to have the opportunity to conduct observations from this mountain.

\facility{Keck:II (NIRSPEC)}

\bibliography{refs}

\begin{deluxetable*}{llllrrrrlrrlrrl}
\tablecaption{Astrometric and RV Data \label{tab:kinematicdata}}
\tablehead{
\colhead{Name} &
\colhead{$\alpha$} &
\colhead{$\delta$} &
\colhead{Ref.} &
\colhead{$\mu_{\alpha *}$} &
\colhead{$\sigma_{\mu_{\alpha *}}$} &
\colhead{$\mu_{\delta}$} &
\colhead{$\sigma_{\mu_{\delta}}$} &
\colhead{Ref.} &
\colhead{$\pi$} &
\colhead{$\sigma_{\pi}$} &
\colhead{Ref.} &
\colhead{RV} &
\colhead{$\sigma_{RV}$} &
\colhead{Ref.}
\\
\colhead{ } &
\colhead{\degr} &
\colhead{\degr} &
\colhead{ } &
\colhead{mas~yr$^{-1}$} &
\colhead{mas~yr$^{-1}$} &
\colhead{mas~yr$^{-1}$} &
\colhead{mas~yr$^{-1}$} &
\colhead{ } &
\colhead{mas} &
\colhead{mas} &
\colhead{ } &
\colhead{km~s$^{-1}$} &
\colhead{km~s$^{-1}$} &
\colhead{ } 
}
\startdata 
2M0253+32   & 43.499173 & 32.110363 &  (1)  & 87 & 10 & -96 & 10 &  (9)  & 17.7 & 2.5 &  (9)  & $-$35.114 & 1.501 &   (3) \\
  &    &    &    & 89.1 & 7.2 & $-$98.3 & 8.5 &  (10)  & 20.22 & 0.18 &  (8)  &    &    &  \\
   & & & & 95.9 & 1.1 & $-$96.9 & 1.1 & (5) & 21.3 & 1.0 & (5) & & & \\
  &    &    &    & 92.49 & 0.38 & $-$100.23 & 0.26 &  (8)  &    &    &    &    &    &  \\
  &    &    &    & {\bf 92.84} & {\bf 0.36} & {\bf $-$100.05} & {\bf 0.25} & & {\bf 20.24} & {\bf 0.18} & & {\bf $-$35.11} & {\bf 1.50} & \\
2M0534$-$06   & 83.566445 & $-$6.52772 &  (1)  & 2 & 12 & -7 & 12 &  (9)  &    &    &    & 28.635 & 2.938 &   (3) \\
  &    &    &    & 2.2 & 19.7 & $-$6.9 & 20.8 &  (10)  &    &    &    &    &    &  \\
  &    &    &    & {\bf 2.05} & {\bf 10.25} & {\bf -6.98} & {\bf 10.39} & & & & & {\bf 28.64} & {\bf 2.94} & \\
2M1935$-$28   & 293.983154 & $-$28.776211 &  (1)  & 34 & 12 & -58 & 12 & (9) & 17.69 & 0.49 &  (8)  & $-$7.736 & 1.278 &   (3) \\
 & & & & 27.3 & 0.9 & $-$61.6 & 1.1 & (5) & 14.2 & 1.2 & (5) & $-$5.08 & 3.48 & (19) \\
  &    &    &    & 27.2 & 4.8 & $-$56.6 & 5.1 &  (10)  &    &    &    &    &    &  \\
  &    &    &    & 26.40 & 0.60 & $-$62.09 & 0.54 &  (8)  &    &    &    &    &    &  \\
  &    &    &    & {\bf 26.70} & {\bf 0.50} & {\bf $-$61.94} & {\bf 0.48} & & {\bf 17.19} & {\bf 0.45} & & {\bf $-$7.42} & {\bf 1.20} & \\
2M0027+05  & 6.924889 & 5.0616 &  (1)  & 10.5 & 0.4 & $-$0.8 & 0.3 &  (2)  & 13.8 & 1.6 &  (2)  & 6.788 & 1.541 &   (3) \\
  &    &    &    & 16.1 & 1.1 & $-$0.33 & 1.2 & (5) & 10.4 & 0.8 & (5)  &    &    &  \\ 
  &    &    &    & {\bf 11.15} & {\bf 0.38} & {\bf $-$0.77} & {\bf 0.29} &  & {\bf 11.08} & {\bf 0.72} &  & {\bf 6.79} & {\bf 1.54} & \\
2M0241$-$03   & 40.297996 & $-$3.449661 &  (1)  & 73.7 & 1 & $-$24.2 & 1.9 &  (9)  & 26.7 & 3.3 &  (9)  & 10.221 & 2.12 &   (3) \\
  &    &    &    & 93.43 & 17 & $-$19.87 & 13.4 &  (14)  & 21.4 & 2.6 &  (4)  &  6.34  &  7.98  & (9) \\
   & & & & 69.6 & 0.5 & $-$25.1 & 0.6 & (5) & 18.5 & 2.1 & (5) & & & \\ 
  &    &    &    & 84 & 11.7 & $-$22.4 & 8.6 &  (4)  &    &    &    &    &    &  \\
  &    &    &    & 76.6 & 12.8 & $-$24.5 & 9.7 &  (10)  &    &    &    &    &    &  \\
  &    &    &    & {\bf 70.46} & {\bf 0.45} & {\bf $-$25.00} & {\bf 0.57} &  & {\bf 21.03} & {\bf 1.46} &  & {\bf 9.97} & {\bf 2.05} & \\
2M0117$-$34   & 19.447838 & $-$34.057171 &  (1)  & 84 & 15 & $-$45 & 8 &  (9)  & 25.56 & 0.71 &  (8)  & 3.258 & 1.351 &   (3) \\
 & & & & 111.5 & 2.1 & $-$ 52.4 & 3.8 & (5) & 26.1 & 1.9 & (5) & & & \\
  &    &    &    & 103.14 & 13.98 & $-$39.7 & 7 &  (14)  &    &    &    &    &    &  \\
  &    &    &    & 102.6 & 6.9 & $-$42.5 & 5.6 &  (10)  &    &    &    &    &    &  \\
  &    &    &    & 108.19 & 0.92 & $-$59.99 & 1.27 &  (8)  &    &    &    &    &    &  \\
  &    &    &    & {\bf 108.54} & {\bf 0.834} & {\bf $-$55.88} & {\bf 1.21} &  & {\bf 25.63} & {\bf 0.67} &  & {\bf 3.26} & {\bf 1.35} & \\
2M0045+16  & 11.339304 & 16.579082 &  (1)  & 355 & 10 & $-$40 & 10 &  (4)  & 57.3 & 2 &  (4)  & 3.287 & 1.333 &   (3) \\
 &  &  &  & 354.4 & 2.2 & $-$51.1 & 2 & (5) & 65.9 & 1.3 & (5) & 3.29 & 0.17 & (6) \\
  &    &    &    & 385 & 17 & $-$26 & 12 & (7) & 65.02 & 0.23 &  (8)  & 3.16 & 0.83 &  (9) \\
  &    &    &    & 374.9 & 8.5 & $-$27.7 & 8.4 &  (10)  &  &  &  &  &  & \\
  &    &    &    & 358.92 & 0.40 & $-$48.07 & 0.24 &  (8)  &    &    &    &    &    &  \\
  &    &    &    & {\bf 358.82} & {\bf 0.39} & {\bf $-$48.08} & {\bf 0.24} &  & {\bf 64.95} & {\bf 0.23} &  & {\bf 3.29} & {\bf 0.17} & \\
2M1551+09   & 237.968246 & 9.687469 &  (1)  & $-$70 & 22 & $-$50 & 22 &  (15)  & 22.1 & 1.5 & (5) & $-$15.389 & 1.451 &   (3) \\
 & & & & $-$62.1 & 0.6 & $-$57.7 & 0.6 & (5) & & & & & & \\  
  &    &    &    & $-$69.4 & 11.1 & $-$55.9 & 11.4 &  (10)  &    &    &    &    &    &  \\
  &    &    &    & {\bf $-$62.09} & {\bf 0.59} & {\bf $-$57.69} & {\bf 0.60} &  & {\bf 22.10} & {\bf 1.50} &  & {\bf $-$15.39} & {\bf 1.45} & \\
2M1615+49   & 243.927302 & 49.889214 &  (1)  & $-$80 & 12 & 18 & 12 &  (9) & 32.0 & 1.0 & (5) & $-$24.018 & 1.697 &   (3) \\
  &    &    &    & $-$23 & 34 & 41.8 & 45.5 &  (16)  &    &    &    & $-$25.59   & 3.18   & (9) \\
   & & & & $-$92.8 & 1.2 & 15.2 & 1.8 & (5) & & & & & & \\ 
  &    &    &    & $-$78.8 & 15.6 & 19.4 & 9.9 &  (10)  &    &    &    &    &    &  \\
  &    &    &    & {\bf $-$92.51} & {\bf 1.19} & {\bf 15.43} & {\bf 1.75} &  & {\bf 32.00} & {\bf 1.00} &  & {\bf $-$24.37} & {\bf 1.50} & \\
2M2154-10   & 328.643928 & $-$10.925234 &  (1)  & 175 & 12 & 9 & 12 &  (9) & 32.6 & 1.0 & (5) & $-$21.361 & 1.715 &   (3) \\
 & & & & 166.8 & 1.7 & 2.2 & 2.2 & (5) & & & & & & \\ 
  &    &    &    & 169.2 & 8.6 & $-$1.6 & 8.8 &  (17)  &    &    &    &  &  & \\
  &    &    &    & {\bf 167.04} & {\bf 1.65} & {\bf 2.19} & {\bf 2.10} &  & {\bf 32.60} & {\bf 1.00} &  & {\bf $-$21.36} & {\bf 1.72} & \\
W0047+68  & 11.751611 & 68.065102 &  (11)  & 387 & 4 & $-$197 & 4 &  (12)  & 82 & 3 &  (12)  & $-$17.094 & 2.732 &   (3) \\
  &    &    &    & 370 & 10 & $-$210 & 10 &  (13)  & 82.3 & 1.8 & (5) & $-$20.0 & 1.4 &   (12) \\
    &    &    &    & 380.7 & 1.1 & $-$204.2 & 1.4 & (5) & & &  &    &    & \\ 
  &    &    &    & 375.3 & 2.9 & $-$212.8 & 9.3 &  (18)  &    &    &    &    &    &  \\
  &    &    &    & {\bf 380.35} & {\bf 0.99} & {\bf $-$203.71} & {\bf 1.30} &  & {\bf 82.22} & {\bf 1.54} &  & {\bf $-$19.40} & {\bf 1.25} & \\
\enddata
\tablecomments{Data sources: (1) \citet{Cutri03} [2MASS], (2) \citet{Dahn02}, (3) This Work, (4) \citet{ZapOso14}, (5) \citet{Liu16}, (6) \citet{Blake10}, (7) \citet{Jameson08}, (8) \citet{Gaia18}, (9) \citet{Faherty16}, (10) \citet{Gagne14a}, (11) \citet{Cutri12} [WISE], (12) \citet{Gizis15}, (13) \citet{Thompson13}, (14) \citet{Casewell08}, (15) \citet{Faherty09}, (16) \citet{Schmidt10}, (17) \citet{Gagne14b}, (18) \citet{Gizis12}, (19) \citet{Shkolnik17}. Values in {\bf bold} are weighted means.}

\end{deluxetable*}

\begin{deluxetable*}{lcrrrrrr}
\tabletypesize{\small}
\tablecaption{\bf Spatial \& Kinematic Properties \label{tab:UVWXYZ}}
\tablehead{ 
\colhead{Object} &
\colhead{Sp. Type} &
\colhead{X} &
\colhead{Y} &
\colhead{Z} &
\colhead{U} &
\colhead{V} &
\colhead{W} \\
\colhead{Name} &
\colhead{(Optical)} &
\colhead{pc} &
\colhead{pc} &
\colhead{pc} &
\colhead{km s$^{-1}$} &
\colhead{km s$^{-1}$} &
\colhead{km s$^{-1}$} }
\startdata
2M0253+32	&	M7$\beta$	&	$-$39.53	$\pm$	0.35	&	21.80	$\pm$	0.19	&	$-$20.07	$\pm$	0.18	&	16.76	$\pm$	1.21	&	$-$44.08	$\pm$	0.71	&	 5.54	$\pm$	0.62	\\
2M1935$-$28	&	M9$\beta$	&	53.14	$\pm$	1.41	&	10.05	$\pm$	0.27	&	$-$21.54	$\pm$	0.57	&	$-$8.89	$\pm$	1.10	&	$-$15.47	$\pm$	0.45	&	$-$9.11	$\pm$	0.56	\\
2M0027+05	&	M9.5$\beta$	&	$-$18.30	$\pm$	1.20	&	45.42	$\pm$	2.97	&	$-$76.27	$\pm$	4.99	&	$-$5.34	$\pm$	0.43	&	 0.77	$\pm$	0.80	&	$-$6.33	$\pm$	1.30	\\
2M0241$-$03	&	L0$\gamma$	&	$-$27.54	$\pm$	1.94	&	 2.10        $\pm$	0.15	&	$-$38.98	$\pm$	2.75	&	$-$12.87	$\pm$	1.29	&	$-$14.32	$\pm$	1.05	&	$-$3.90	$\pm$	1.70	\\
2M0117$-$34	&	L1$\beta$        &	$-$0.50	$\pm$	0.01	&	$-$6.05	$\pm$	0.16	&	$-$38.58	$\pm$	1.01	&	$-$10.09	$\pm$	0.32	&	$-$20.48	$\pm$	0.60	&	 0.05	$\pm$	1.34	\\
2M0045+16	&	L2$\gamma$	&	$-$5.45	$\pm$	0.02	&	 9.14	$\pm$	0.03	&	$-$11.13	$\pm$	0.04	&	$-$21.91	$\pm$	0.10	&	$-$14.13	$\pm$	0.11	&	$-$5.42	$\pm$	0.12	\\
2M2154$-$10	&	L4$\beta$        &	15.10	$\pm$	0.47	&	15.35	$\pm$	0.47	&	$-$21.88	$\pm$	0.67	&	$-$29.46	$\pm$	1.05	&	$-$13.39	$\pm$	0.90	&	 0.24	$\pm$	1.32	\\
2M1551+09	&	L4$\gamma$	&	30.79	$\pm$	2.12	&	10.87	$\pm$	0.75	&	 31.64	$\pm$	2.18	&	$-$8.46	$\pm$	1.00	&	$-$21.36	$\pm$	1.27	&	$-$6.55	$\pm$	1.05	\\
2M1615+49	&	L4$\gamma$	&	 4.72        $\pm$      0.15	&	21.55	$\pm$	0.68	&	 22.18	$\pm$	0.70	&	$-$10.40	$\pm$	0.40	&	$-$24.77	$\pm$	1.07	&	$-$8.09	$\pm$	1.11	\\
W0047+68	&	L7$\gamma$	&	$-$6.51	$\pm$	0.12	&	10.22	$\pm$	0.19	&	 1.10	$\pm$	0.02	&	$-$8.53	$\pm$	0.76	&	$-$27.04	$\pm$	1.07	&	$-$13.83	$\pm$	0.26 \\
\enddata
\end{deluxetable*}

\clearpage

\begin{deluxetable*}{lll|cccccc}
\tabletypesize{\small}
\tablewidth{0pt} 
\tablecaption{\bf Membership Results
\label{tab:memberships} }
\tablehead{ 
\colhead{Object} &
\colhead{Sp. Type} &
\colhead{Final} &
\colhead{LACEwING} &
\colhead{BANYAN I} &
\colhead{BANYAN II} &
\colhead{CONVERGE\tablenotemark{a}} &
\colhead{BANYAN $\Sigma$} \\
\colhead{Name} &
\colhead{(Optical)} &
\colhead{Membership} &
\colhead{$$} &
\colhead{$$} &
\colhead{$$} &
\colhead{$$} &
\colhead{$$}}
\startdata
2M0253$+$32 & M7$\beta$   & None & None & Field & Field & ($\beta$ Pic-92)\tablenotemark{a} & Field \\
2M0534$-$06 & M8$\gamma$  & None & None & Field & Argus-94 & (AB Dor-100) & Field \\
2M1935$-$28 & M9$\gamma$  & $\beta$ Pic & $\beta$ Pic-63 & $\beta$ Pic-100 & $\beta$ Pic-100 & $\beta$ Pic-98 & $\beta$ Pic-99 \\
2M0027$+$05 & M9.5$\beta$ & None & None & Field & Field & (Car-Near-100) & Field \\
2M0241$-$03 & L0$\gamma$  & None & Field & Field & Field & Field & Field \\
2M0117$-$34 & L1$\beta$   & Tuc-Hor & Tuc-Hor-96 & Tuc-Hor-100 & Tuc-Hor-100 & Tuc-Hor-99 & Tuc-Hor-100 \\
2M0045$+$16 & L2$\beta$   & Argus & Argus-98 & Argus-100 & Argus-100 & Field\tablenotemark{b} & Argus-100 \\
2M1551$+$09 & L4$\gamma$  & None & None & Field & Field & Field & Field \\
2M1615$+$49 & L4$\gamma$  & None & AB Dor-25 & Field & (AB Dor-37) & (Tuc-Hor-86) & Field \\
2M2154$-$10 & L4$\beta$   & Carina-Near & Carina-Near-53 & Field\tablenotemark{b} & Field & Field & Car-Near-89 \\
W0047$+$68 & L7$\gamma$  & AB Dor & AB Dor-100 & AB Dor-100 & AB Dor-100 & AB Dor-85 & AB Dor-100 \\
\enddata
\tablecomments{The quoted membership probability is the highest membership probability for the most commonly identified moving group, considering every permutation of kinematic data. Probabilities in parentheses are below the quality threshold: LACEwING: 20\%; BANYAN codes, 90\%; Convergence code, 80\%. See discussion in Section \ref{sec:NYMGs}.}

\tablenotemark{a}{Values in parentheses are inconsistent with actual membership. Either the probability is too low for that particular code, or (particularly for the Convergence code) the predicted distance, space position, or radial velocities are inconsistent with membership or actual measurements.}
\tablenotetext{b}{The Convergence Code and BANYAN $\Sigma$ do not consider membership in Argus. Only LACEwING, the Convergence code, and BANYAN $\Sigma$ consider membership in Carina-Near.}

\end{deluxetable*}


\begin{deluxetable*}{cccccccc}
\tabletypesize{\scriptsize}
\tablewidth{0pt}
\tablecolumns{8}
\tablecaption{Spectral Line Measurements}
\label{tab:phew_tab}
\tablehead{ 
\colhead{} & \colhead{} & \multicolumn{3}{c}{1.2436 $\mu$m} & \multicolumn{3}{c}{1.2525 $\mu$m}   \\
\cline{3-8} \\
\colhead{Object Name} &  \colhead{Optical SpT} &
\colhead{EW (\AA)} & \colhead{FWHM (km/s)} & \colhead{SNR} & \colhead{EW (\AA)} & \colhead{FWHM (km/s)}  & \colhead{SNR} 
}
\startdata 
\cutinhead{Suspected Young Objects}
2M0253+32 & 	M7 $\beta$	 &  2.50 $\pm$ 0.60 & 38.60  $\pm$ 3.16  & 21.86 & 2.39 $\pm$ 0.57 & 45.51  $\pm$ 5.00   & 21.86 \\
2M0534-06 & 	M8 $\gamma$	 &  2.80 $\pm$ 2.30 & 86.84  $\pm$ 62.72 & 3.85  & 1.94 $\pm$ 1.10 & 74.25  $\pm$ 12.55  & 3.92  \\
2M1935$-$28 & 	M9 $\gamma$	 &  2.31 $\pm$ 0.40 & 36.82  $\pm$ 2.68  & 23.46 & 3.11 $\pm$ 0.17 & 50.86  $\pm$ 4.46   & 24.10 \\
2M0027+05 & 	M9.5 $\beta$ &  3.68 $\pm$ 2.35 & 101.32 $\pm$ 9.79  & 3.75  & 3.60 $\pm$ 1.24 & 69.46  $\pm$ 12.07  & 4.01  \\
2M0241$-$03 & 	L0 $\gamma$	 &  3.41 $\pm$ 1.20 & 137.50 $\pm$ 3.68  & 5.42  & 4.18 $\pm$ 0.60 & 134.13 $\pm$ 7.84   & 5.64  \\
2M0117$-$34 & 	L1 $\beta$	 &  2.64 $\pm$ 1.01 & 62.72  $\pm$ 6.22  & 7.74  & 4.54 $\pm$ 0.31 & 86.23  $\pm$ 9.52   & 8.01  \\
2M0045+16 & 	L2 $\beta$	 &  2.47 $\pm$ 0.16 & 79.25  $\pm$ 5.41  & 34.41 & 4.66 $\pm$ 0.07 & 99.32  $\pm$ 7.43   & 36.18 \\
2M2154$-$10 & 	L4 $\beta$	 &  3.40 $\pm$ 2.74 & 65.13  $\pm$ 6.85  & 3.97  & 5.27 $\pm$ 3.21 & 93.41  $\pm$ 30.39  & 4.13  \\
2M1551+09 & 	L4 $\gamma$	 &  2.81 $\pm$ 1.40 & 55.48  $\pm$ 2.65  & 5.05  & 4.30 $\pm$ 0.60 & 76.65  $\pm$ 9.79   & 5.31  \\
2M1615+49 & 	L4 $\gamma$	 &  3.80 $\pm$ 1.56 & 101.32 $\pm$ 12.70 & 3.89  & 3.49 $\pm$ 1.20 & 146.11 $\pm$ 25.36  & 3.99  \\
W0047+68 & 	L7 $\gamma$	 &  1.54 $\pm$ 0.81 & 48.25  $\pm$ 13.04 & 6.25  & 2.40 $\pm$ 0.92 & 62.28  $\pm$ 9.81   & 6.28  \\
\cutinhead{Field M \& L Dwarfs}
Wolf 359\tablenotemark{a}          &  M6     &  5.60    $\pm$ 0.60   & 65.00  $\pm$ 7.00  &  1.00   & 5.20   $\pm$  0.50  & 64.00   $\pm$ 6.00  &  1.00  \\
2MASS J0140+2701\tablenotemark{a}  &  M9     &  9       $\pm$ 0.9    & 78.00  $\pm$ 8.00  &  76.72  & 7.5    $\pm$  0.7   & 110.00  $\pm$ 11.00 & 81.33 \\
2MASS J0345+2540\tablenotemark{a}  &  L0     &  11.50   $\pm$ 1.20   & 220.00 $\pm$ 22.00 &  23.43  & 9.30   $\pm$  0.90  & 180.00  $\pm$ 18.00 & 25.48 \\
2MASS J0746+2000\tablenotemark{a}  &  L0.5   &  14.1    $\pm$ 1.4    & 230.00 $\pm$ 23.00 &  64.34  & 11.5   $\pm$  1.1   & 210.00  $\pm$ 21.00 & 65.76 \\
2MASS J0208+2542\tablenotemark{b}  &  L1     &  10.1    $\pm$ 0.22   & 81.19  $\pm$ 4.54  &  34.42  & 15.48  $\pm$  0.25  & 104.68  $\pm$ 3.59  & 36.11  \\
2MASS J1658+7027\tablenotemark{b}  &  L1     &  10.22   $\pm$ 0.28   & 74.42  $\pm$ 1.58  &  22.25  & 15.08  $\pm$  0.32  & 95.23   $\pm$ 10.20 & 22.72  \\
2MASS J2057$-$0252\tablenotemark{b}  &  L1.5   &  10.91   $\pm$ 0.2    & 103.72 $\pm$ 3.38  &  38.86  & 15.2   $\pm$  0.23  & 153.62  $\pm$ 7.45  & 40.56  \\
2MASS J2130$-$0845\tablenotemark{b}  &  L1.5   &  11.88   $\pm$ 1.04   & 88.58 $\pm$ 8.46 &  7.04   & 17.36   $\pm$  1.28  & 117.20 $\pm$ 3.61 & 6.97   \\
Kelu-1AB\tablenotemark{a}          &  L2     &  14.10   $\pm$ 1.40   & 320.00 $\pm$ 31.00 &  1.00   & 14.10   $\pm$  1.40  & 320.00  $\pm$  32.00 & 1.00   \\
2MASS J2104$-$1037\tablenotemark{b}  &  L2.5   &  11.05   $\pm$ 0.23   & 93.60 $\pm$ 1.52 &  32.98  & 16.37   $\pm$  0.26  & 124.87 $\pm$ 10.16 & 33.36  \\
2MASS J1506+1321\tablenotemark{b}  &  L3     &  11.34   $\pm$ 0.06   & 104.01 $\pm$ 5.80 &  39.21  & 16.95   $\pm$  0.06  & 139.84 $\pm$ 5.72 & 40.19  \\
2MASSW J0036+1821\tablenotemark{b} &  L3.5   &  11.40   $\pm$ 1.10   & 290.00 $\pm$ 29.00 &  62.75  & 14.40   $\pm$  1.40  & 240.00  $\pm$  24.00 & 66.64  \\
GD165B\tablenotemark{a}            &  L4     &  14.00   $\pm$ 1.40   & 150.00 $\pm$ 15.00 &  4.58   & 12.60   $\pm$  1.30  & 230.00  $\pm$  23.00 & 4.83   \\
2MASS J1821+1414\tablenotemark{b}  &  L4.5   &  9.69    $\pm$ 0.49   & 89.61 $\pm$ 2.89 &  19.84  & 15.62   $\pm$  0.54  & 128.95 $\pm$ 7.28 & 20.56  \\
2MASS J2224$-$0158\tablenotemark{b}  &  L4.5   &  12.81   $\pm$ 0.45   & 83.53 $\pm$ 1.36 &  19.44  & 17.88   $\pm$  0.54  & 121.67 $\pm$ 3.74 & 19.50  \\
2MASS J0835+1953\tablenotemark{b}  &  L5     &  4.18    $\pm$ 0.03   & 67.55 $\pm$ 29.93  &  4.54   & 4.18    $\pm$  0.03  & 150.54 $\pm$ 3.15 & 4.34   \\
2MASS J1507$-$1627\tablenotemark{a}  &  L5     &  14.70   $\pm$ 1.50   &  270 $\pm$ 27 &  1.00   & 10.00   $\pm$  1.00  & 240  $\pm$  24 &  1.00  \\
2MASS J0103+1935\tablenotemark{b}  &  L6     &  7.46    $\pm$ 1.46   & 176.10 $\pm$ 25.10  &  4.62   & 15.49   $\pm$  2.63  & 107.18 $\pm$ 11.23 & 4.50   \\
2MASS J1010$-$0406\tablenotemark{b}  &  L6     &  2.60    $\pm$ 0.08   & 96.49 $\pm$ 13.52 & 7.06   & 4.16    $\pm$  0.03  & 126.51 $\pm$ 5.11 & 6.81   \\
DENIS J0205$-$1159\tablenotemark{a}  &  L7     &  8.20    $\pm$ 1.20   & 390.00 $\pm$ 39.00 &  7.56   & 8.60    $\pm$  1.30  & 290.00  $\pm$  29.00 & 7.55   \\
\enddata    
\tablecomments{Objects are grouped by spectral type and then listed in order of right ascension.}
\tablenotetext{a}{Originally published in \citet{McLean07}.}
\tablenotetext{b}{Originally published in Alam et. al (in prep).}
\end{deluxetable*}

\end{document}